\documentclass[fleqn,10pt]{wlscirep}
\usepackage[utf8]{inputenc}
\usepackage[T1]{fontenc}
\usepackage{nameref}
\usepackage{soul}
\usepackage{pifont}
\usepackage{xcolor} 

\usepackage[final]{changes}

\title{X-ray2CTPA: Leveraging Diffusion Models to Enhance Pulmonary Embolism Classification}
\author[1,*]{Noa Cahan}
\author[2]{Eyal Klang}
\author[3]{Galit Aviram}
\author[4]{Yiftach Barash}
\author[4]{Eli Konen}
\author[1]{Raja Giryes}
\author[1,5]{Hayit Greenspan}
\affil[1]{Faculty of Engineering, Tel Aviv University, Tel-Aviv, Israel}
\affil[2]{Division of Data-Driven and Digital Medicine, Department of Medicine, Icahn School of Medicine at Mount Sinai, New York, NY 10029, United States.}
\affil[3]{Department of Radiology, Tel-Aviv Sourasky Medical Center and Tel Aviv University School of Medicine, Israel.}
\affil[4]{Department of Diagnostic Imaging, Sheba Medical Center, Ramat Gan, Israel affiliated with the Tel Aviv University, Tel Aviv, Israel}
\affil[5]{Department of Radiology, Icahn School of Medicine, Mount Sinai,NY}
\affil[*]{Corresponding author. E-mail: noa.cahan@gmail.com}

\begin{abstract}

Chest X-rays or chest radiography (CXR), commonly used for medical diagnostics, typically enables limited imaging compared to computed tomography (CT) scans, which offer more detailed and accurate three-dimensional data, particularly contrast-enhanced scans like CT Pulmonary Angiography (CTPA). However, CT scans entail higher costs, greater radiation exposure, and are less accessible than CXRs. In this work, we explore cross-modal translation from a 2D low contrast-resolution X-ray input to a 3D high contrast and spatial-resolution CTPA scan. Driven by recent advances in generative AI, we introduce a novel diffusion-based approach to this task. We employ the synthesized 3D images in a classification framework and show improved AUC in a Pulmonary Embolism (PE) categorization task, using  the initial CXR input. Furthermore, we evaluate the model’s performance using quantitative metrics, ensuring diagnostic relevance of the generated images. 
The proposed method is generalizable and capable of performing additional cross-modality translations in medical imaging. It may pave the way for more accessible and cost-effective advanced diagnostic tools. The code for this project is available: \url{https://github.com/NoaCahan/X-ray2CTPA}.

\end{abstract}


\begin{document}

\flushbottom
\maketitle
%
%
\thispagestyle{empty}

\section*{Introduction}




CXRs are pivotal in emergency diagnostics due to their speed and accessibility, offering a cost-effective, non-invasive solution \cite{xray}. However, their 2D nature and lower soft tissue contrast limit their diagnostic ability compared to CT scans, which provide comprehensive 3D views of the various tissues including the lungs, bone and soft tissue, albeit at higher costs and radiation exposure \cite{ct, ct2}. CTPA scans, a specialized form of CT with  fast contrast material injection, further enhance blood vessels but may be less suitable for certain groups, including those with severe contrast allergies, or renal failure.
In this work, we wish to demonstrate the use of advanced generative AI schemes for cross-model translation and to use the generated data to enhance a classification task. In our main focus scenario, we build a system to translate 2D chest X-ray images into 3D CTPA scans. We explore the quality of the generated scans and their promise for earlier detection of findings in the initial CXR data, specifically focusing on PE classification. Finally, We show the generalization  of the developed cross-model translation scheme using the LIDC dataset, demonstrating possible transition from CXR to synthetic thoracic CT images.

We address the classification of PE as present or absent, from CXR imaging as a case study in order to demonstrate the potential of transferring CXR to CT. 
PE is a life-threatening condition where a blood clot that develops in a limb vein travels to the pulmonary arteries where it suddenly blocks the blood flow. The reference standard for PE diagnosis is CTPA – a 3D lung CT with contrast injection. However, the use of CT is associated with exposure to ionizing radiation, and iodinated contrast, requires expensive scanners and qualified staff to operate them, and radiologists to interpret. The number of CTPAs performed continues to rise, with only 5-15\% of them being positive for PE diagnosis \cite{positive_pe}.
Therefore, there is a need in the modern health care system to develop a simple, easy, fast and objective tool for diagnosing this common but life-threatening disease of PE, based on low-radiation-low-cost data, which is routinely collected for the investigation of patients with cardio-respiratory complaints.
In the emergency department, almost every patient with cardiac or respiratory issues undergoes a CXR, despite its lower tissue resolution and limited information for PE diagnosis. This underscores the opportunity to enhance the utility of CXR imaging in PE detection through innovative approaches.

Given the substantial drawbacks associated with CT - such as high costs, significant radiation exposure, and logistic bottlenecks - there is a critical need to leverage more accessible data sources like CXRs. However, CXRs alone often lack sufficient detail for definitive diagnosis, underscoring the opportunity to introduce advanced AI-based methods. By harnessing generative and discriminative deep learning schemes, we aim to supplement the limited information in CXRs with high-fidelity, synthesized 3D CTPA representations that could ultimately improve diagnostic outcomes and reduce unnecessary scans.



Recent years have witnessed remarkable advancements in Generative AI, a field dedicated to empowering machines with the ability to create novel, realistic data using frameworks such as Generative Adversarial Networks (GANs) and diffusion models. From crafting stunning photorealistic images to composing original music and even translating languages fluently \cite{hendy2023good,smith}, generative models have demonstrated their versatility and potential across diverse domains. However, in the field of healthcare and medical imaging, the integration of cross-modal generative AI, especially in 3D imaging, is still in an early stage and confronts significant challenges \cite{ai_in_medical}. Specifically, the synthesis of high-resolution CTPA scans from an extremely limited number of 2D scan slices  and furthermore from  a single 2D image is a yet unaddressed challenge.

A limited number of recent works have shown possible transition from digitally reconstructed Radiographs (DRRs) to CT volumes \cite{x2ct,mednerf}. Here, the input is synthetic CXR data that is generated from a CT's 2D projection. The use of this synthetic approach simplifies the problem and makes these works 
somewhat remote from the true clinical setting. The reason is that DRRs inherently contain more information compared to actual CXR scans, and they are inherently aligned with the CT data from which they are extracted. 
This intrinsic registration and enhanced information content potentially ease the conversion task from DRRs to CT, when contrasted with using real CXRs. 
Furthermore, we are not aware of any existing work that has demonstrated the potential of using the generated synthesized scans for a specific diagnostic application.  

In general, CXR to CT conversion is an under-explored area of research and there are not many works that attempt to solve this problem. X2CT-GAN \cite{x2ct} and it's extension - XprospeCT\cite{paulson2024xprospect}, explores the use of GANs for reconstructing 3D CT volumes from biplanar 2D CXRs. MedNeRF \cite{mednerf} introduces a deep learning model for reconstructing CT projections from as few as a single-view CXR. This approach utilizes a neural radiance field (NeRF) architecture to learn a continuous representation of CT scans. The model effectively disentangles the shape and volumetric depth of surface and internal anatomical structures from 2D images. For both papers, the CXRs used are synthetic constructions from CTs called DRR, which are similar but not identical to real CXRs. This is due to the lack of a publicly available dataset of paired CXR and CT.
We argue that the task of converting DRRs to CT scans is less complex than our task because DRRs inherently possess more information than CXRs. Also, DRRs are naturally aligned and registered with CTPAs due to the way they are constructed.

A concurrent work, X-Diffusion \cite{x_diffusion} is perhaps the most related to our approach. It generates 3D MRI volumes from either a single slice of the MRI or from a registered Dual-energy X-ray Absorptiometry (DXA) which is a single image data modality that is similar to X-ray but includes other non-bony information such as tissue mass. Unlike their framework, the challenges in our problem setting are heightened by the limited size of our dataset, the extremely low resolution in the CXR scans, and the fact that, in our case, the two data modalities are completely distinct and unaligned or unregistered. To the best of our knowledge, our model is the first one to convert matching 2D X-rays to 3D CT. 

Image generation is one of the primary objectives of diffusion models, which has been widely applied in a variety of styles, including generating synthetic 3D medical images \cite{kazerouni2022diffusion}. Medical Diffusion \cite{medical_diffusion} excelled in creating high-quality medical 3D data in an unconditional setting. There are also many 3D conditional models. We list here a few but there are many more: DISPR \cite{dispr} involves reconstructing 3D cells from 2D cell images and their corresponding segmentation mask, Med-DDPM model \cite{med-ddpm} is a conditional diffusion model for semantic 3D medical image synthesis.
Make-A-Volume \cite{makeavolume} leverages latent diffusion models for cross-modality 3D brain MRI synthesis and GenerateCT \cite{generatect} introduced a novel approach that combines transformers and diffusion models for generating CT volumes conditioned on free-form medical text prompts. 

Our work represents a departure from existing works, where conditioning often relies on text prompts trained on extensive datasets or involves adding directly correlated information, such as segmentation masks or structural guides, to the generated image. In contrast, in our research the conditioning is on a different modality altogether. 

\begin{figure}[t]%
\centering
\begin{tabular}{c} 
\includegraphics[scale=0.5]{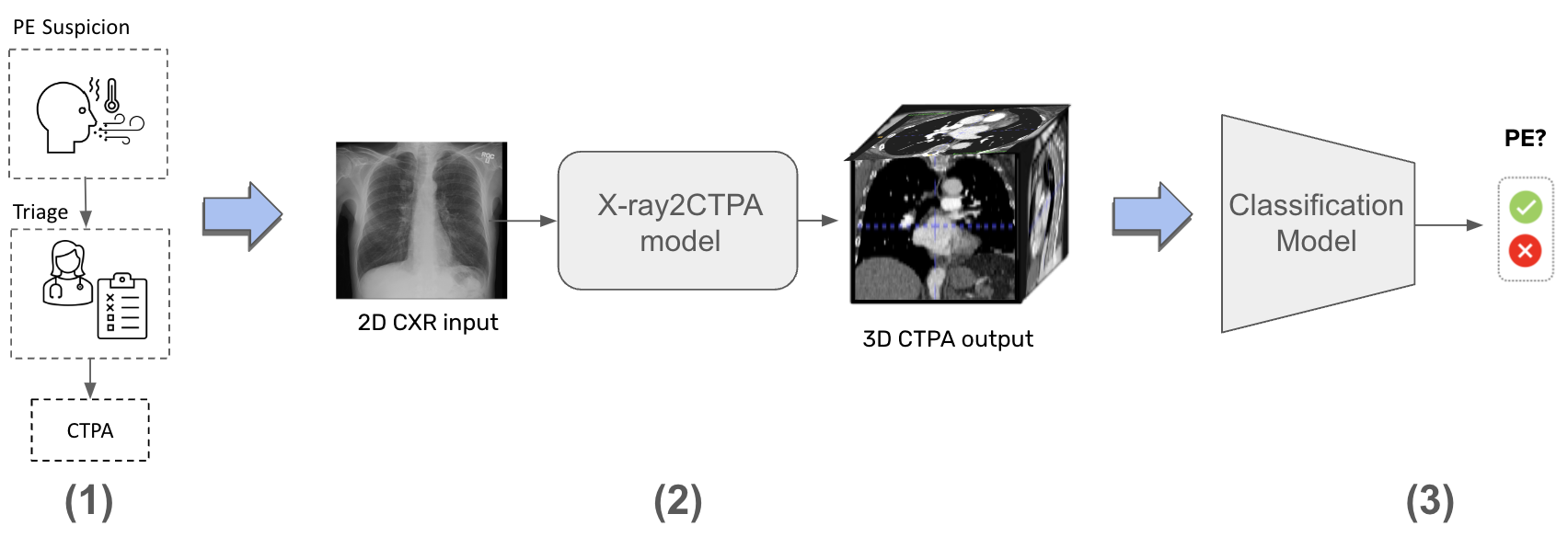}
\end{tabular}
\caption{The outline of our study. The first stage was to establish the cohort and collect the dataset. Our pipeline for PE classification from CXRs consisted of collecting patients who underwent both chest X-rays and CTPA scans. The PE diagnosis label was collected from the CTPA scan's radiologist report. At a second stage, once the X-ray2CTPA diffusion model is trained, the corresponding CTPA scans are generated from their matching CXRs. Finally, the generated data is used for training and testing a PE classifier for PE diagnosis. (Icon images from Microsoft PowerPoint 2023.)}\label{fig:outline}
\end{figure}

Bridging this gap, our study collects a unique dataset of 900 patients who underwent both a 3D CTPA and a 2D CXR within 24 hours, and were suspected to have PE. The CTPAs and the CXRs are paired so that each pair belongs to the same patient. 
The PE diagnosis label (positive or negative), was collected from the CTPA scan's radiologist report of the CTPA.
We use the cross-modality paired scans to train the X-ray2CTPA diffusion model (details in the Methods Section). In a following stage, once the X-ray2CTPA diffusion model is trained, synthesized  "matching" CTPA scans (with or without PE) are generated from an input CXR image and used for training and testing a PE classifier for PE diagnosis. \textbf{Fig. \ref{fig:outline}} presents the study design used for this research. 

It is our hypothesis, that the synthesized data can support PE categorization from the initial CXR imaging phase. 
The generation of CTPAs from CXR images may enhance the specificity of diagnosing PE by CXRs, potentially decreasing the number of patients who require actual CTPA scans, with higher yield (positive PE scans-in higher percentage). Our model can predict the presence of PE on CTPA, leading to a reduction of negative CTPA scans, thus saving the radiation-related health burden and financial health resources currently spent on the performance of redundant negative CTPAs, while ensuring a safe exclusion of PE without the need of CTPA. Note that the generated CTPAs are not clinical substitutes for actual CTPA scans. They can be viewed as  "pseudo-CTPA" information that can provide valuable preliminary data prior to deciding on conducting a comprehensive real CTPA examination. 

Our contributions through this work are as follows: 
\begin{enumerate}
    \item \textbf{Innovative Cross-Modal Translation.} We introduce a novel method, leveraging the power of diffusion models, CLIP Vision encoder, and adversarial guidance to translate 2D CXR images into 3D CTPA scans. This method maintains high fidelity and diagnostic relevance, as demonstrated by both quantitative metrics and improved PE classification performance using the generated data. To the best of our knowledge, this is the first study to utilize real paired CXR and CTPA data for CXR-to-CTPA translation.
    \item \textbf{Potential earlier identification of  PE in CXR.} We demonstrate the use of cross-modality synthetic data generation to increase the performance of PE identification in CXR images. By generating high-quality synthetic 3D images, we show results for improved classification accuracy for PE, increasing the specificity of diagnosing PE.    
    \item \textbf{Generalizability.} The proposed method demonstrates potential for broader applications in medical imaging, paving the way for more accessible advanced diagnostic tools.
\end{enumerate}

\section*{Results} \label{sec:results}

We present below the results of applying our proposed generation approach for PE classification enhancement from CXR as well as various quantitative and qualitative evaluations of the generated samples. Finally, we perform an ablation study in which we explore different model architecture choices.

\begin{figure}[hbt!]
\begin{tabular}{c}  
\includegraphics[scale=0.50]{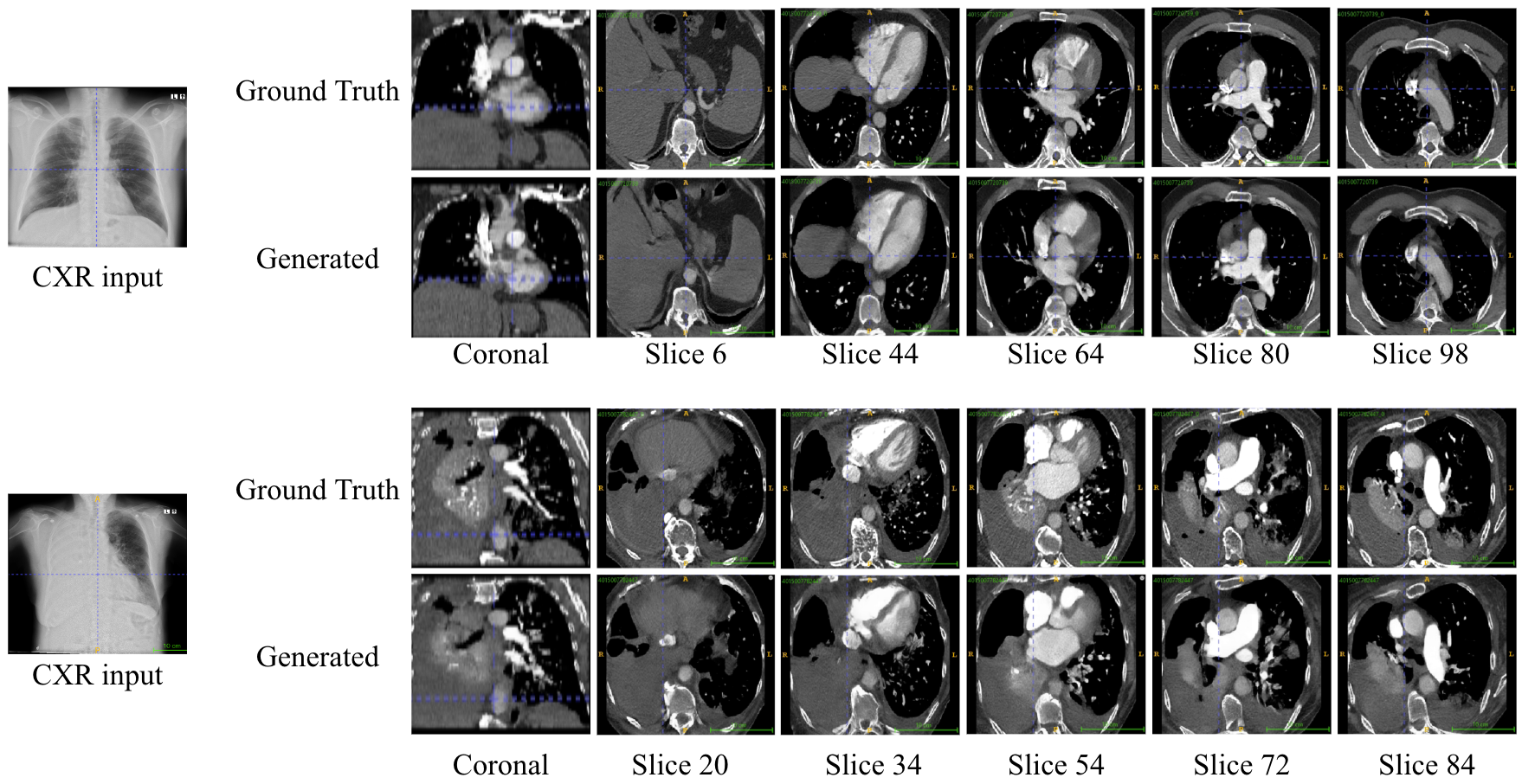} \\
\includegraphics[scale=0.50]{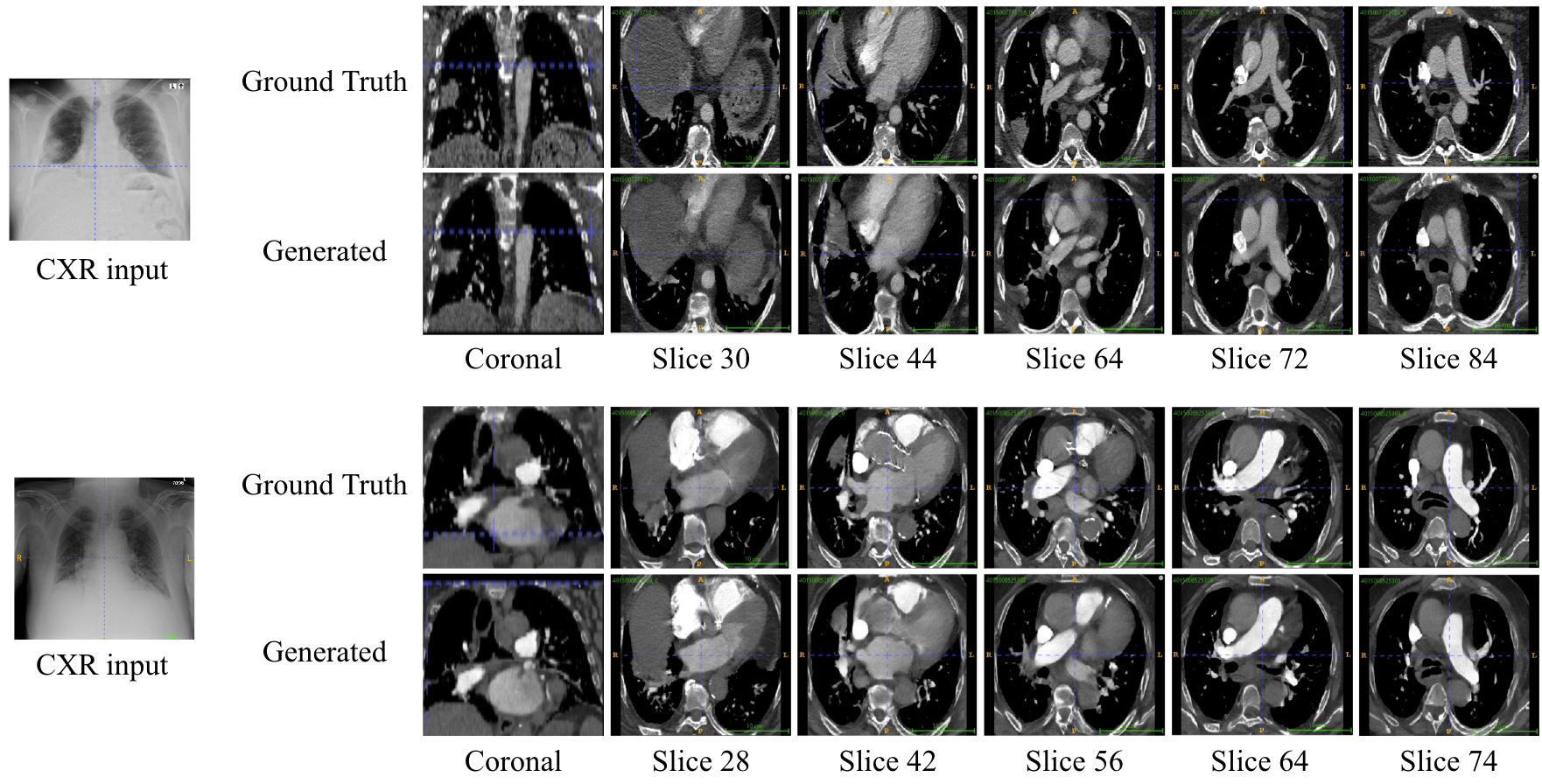} \\
\end{tabular}
\caption{Qualitative results. We show examples of CXR to 3D CTPA Generation with an X-ray2CTPA model. Shown are four randomly selected generated 3D-CTPA samples from the test set, synthesized from a 2D CXR input. We compare slices from the generated 3D-CTPA scan to the ground truth samples. For simplicity, we present five key slices from the whole 3D-scan. The input CXR is presented as well, along with its matching coronal slice from the matching CTPA and generated 3D-CTPA. Note: throughout the paper the axial CT images appear slightly deformed. This is due to the preprocessing of the data and region of interest selection. (We elaborate on this in the Dataset Section.)}
\label{fig:sample_slices}
\end{figure}

\subsection*{Visualizing the Generated Data}

In \textbf{Fig. \ref{fig:sample_slices}} we show visual examples of generated scans. The results demonstrate that the generated scans seem anatomically correct, realistic and consistent between sequential slices. \textbf{Fig. \ref{fig:sample_slices}}. visualizes key slices from four randomly selected generated samples from our held-out test set and the corresponding slices compared with the ground truth scan. As can be seen, the slices are similar but small variations exist compared to the ground truth samples. These differences will be further discussed in the following sections. In addition, the input CXRs for the generated samples are depicted, along with the corresponding coronal example slice from the ground truth and the matching slice from the generated sample. It is evident that general anatomical structure is preserved between the generated sample and ground truth and matches the CXR input.

\begin{figure}[hbt!]
\centering
\begin{tabular}{c}
\includegraphics[scale=0.55]{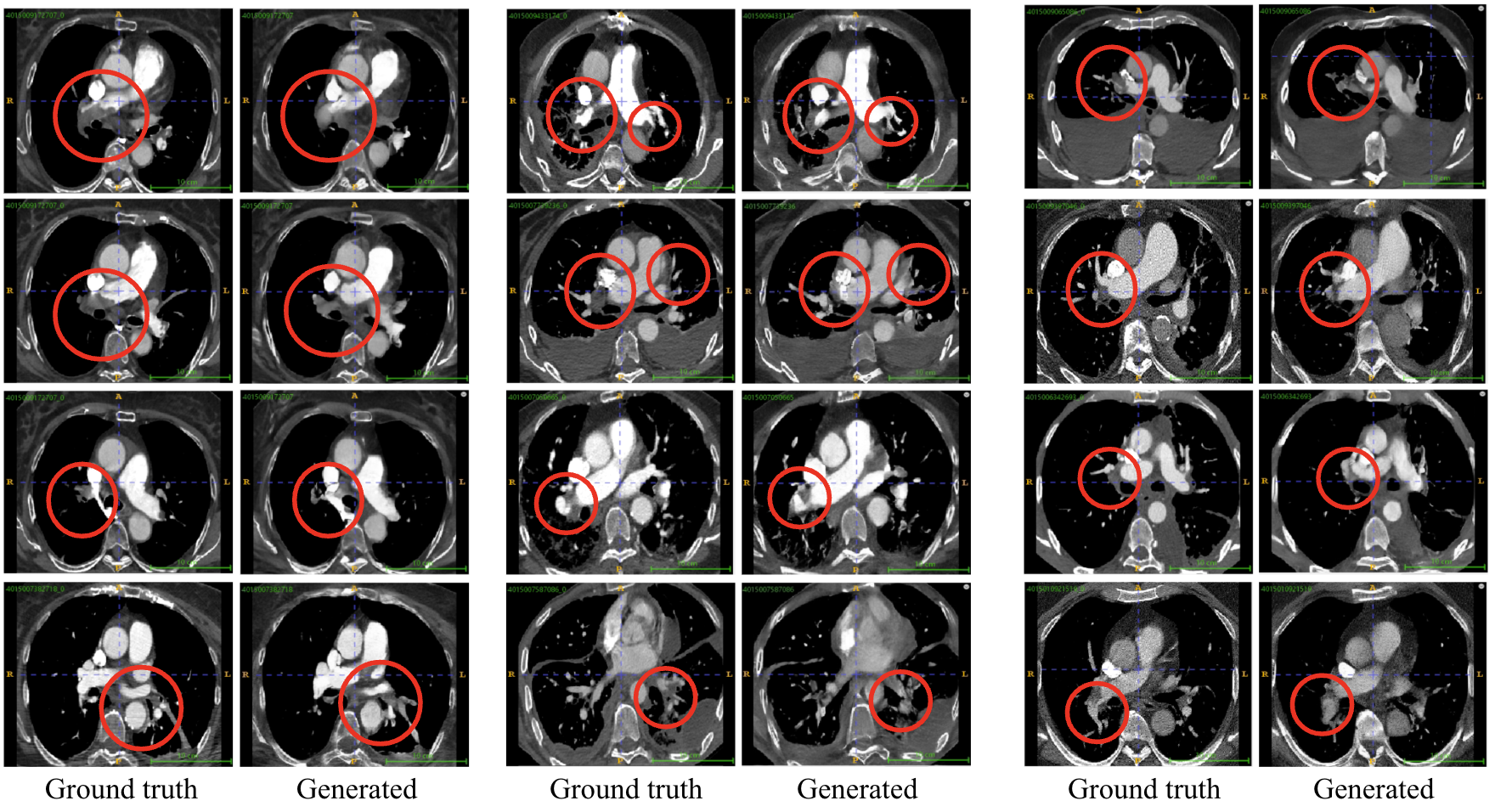}
\end{tabular}
\caption{Visual results of generated PE clots examples. Presented are different 2D slices from generated 3D CTPA samples from our test set, which contain PE clots, compared to their ground truth 2D slices. The PEs are marked with a red circle.}
\label{fig:pe_examples}
\end{figure}

We next compare generated PE examples from the synthesized data with their corresponding ground truth. \textbf{Fig. \ref{fig:pe_examples}} presents the results. As observed, the X-ray2CTPA model successfully generates findings consistent with PE diagnosis. Although some variations in manifestation and location may occur in the generated PE findings compared to the ground truth samples, the overall results are similar. We hypothesize that our model captures the statistical essence of the ground truth, even if the resulting scans are not identical. This hypothesis is supported by the results presented in the following sections. 

\subsection*{Using Generated 3D CTPAs to Enhance PE Classification}
As part of our evaluation, we aimed to assess the generated CTPA's suitability to be used for a classification task. Specifically, we applied the generated data to the task of PE classification. As mentioned in the Introductory section, CTPA is the reference standard for PE diagnosis. PE is usually not noticeable on CXRs due to their non-contrast enhanced and 2D nature. We hypothesis that by training the diffusion model with matching CXR and CTPA pairs, an alignment is constructed between the two modalities that can stir the generated 3D-image so that it can improve the classification results compared to a classification using CXRs.

\textbf{Table \ref{tab:classification_results}} presents the results for this classification task. 
We start with a baseline: The classification of PE using unimodal data, CTPA only and CXR only data. The results present our sixfold cross-validation analysis, by presenting the average AUC, with a 95\% confidence interval, calculated across all folds using the Delong method \cite{delong}.
We set our operating point based on the Youden’s J-Score statistic that maximizes the sum of sensitivity and specificity on the validation set. For the numerical threshold that separates the predicted classes, we utilized the standard definition of the operating point.

\textbf{Table \ref{tab:classification_results}} presents the following findings: First, we note the baseline results, where using CTPA data achieves an AUC of 0.90 [95\% CI: 0.88-0.97], whereas using CXR data yields a significantly lower AUC of 0.691 [95\% CI: 0.54-0.77]. When utilizing the generated data, the model achieves an AUC of 0.803 [95\% CI: 0.72-0.9]. This is a significant improvement over the original CXR data ($\alpha \le 0.05$), approaching the performance benchmark of true CTPA data. Furthermore, the model demonstrates improvements in specificity and NPV compared to the CXR classifier. 
In a clinical context, this outcome suggests a higher number of true negative results, potentially reducing the number of unnecessary CTPA scans performed. These findings underscore that synthetic CTPA data enhances model performance compared to relying solely on CXRs.


\begin{table}[ht!]
\caption{Comparing the average AUC scores (with a 95\% confidence interval) of the PE classification from CXR, CTPA and Generated CTPAs on our test set.} \label{tab:classification_results}
\centering
\begin{tabular}{@{}lcccccc@{}}
\toprule
\multicolumn{1}{c}{Model} & AUC & Accuracy & Specificity & Sensitivity & PPV & NPV\\
\midrule \midrule
\multicolumn{7}{c}{Baselines}\\
\midrule
CTPA only classifier & $0.90$  [$0.88$-$0.97$] $\pm 0.04$ & $84.05$ & $83.86$ & $82.74$ & $69.78$ & $92.32$\\
CXR only classifier & $0.691$ [$0.54$-$0.77$] $\pm 0.06$ & $71.79$ & $70.91$ & $73.91$ & $51.52$ & $86.67$\\
\midrule
\multicolumn{7}{c}{Generated Data}\\
\midrule
Generated CTPA only classifier & $0.80$ [$0.72$-$0.9$] $\pm 0.07$ & $74.0$ & $74.1$ & $73.91$ & $54.8$ & $87.0$\\
\bottomrule
\multicolumn{7}{@{}l@{}}{All AUC results are statistically significant with Delong-ROC test, with $p \leq  0.05$}
\end{tabular}
\end{table}

\subsection*{Generative Model Exploration}
\paragraph*{Quantitative Results.}

To assess the generated sample quality, we conduct quantitative evaluations utilizing various metrics. While these metrics are commonly employed and generally align with human judgment, they are not flawless indicators. There is ongoing exploration of improved metrics in evaluating sample quality, which remains an open challenge. In essence, appraising the visual accuracy of synthetic images, particularly within medical imaging, continues to be a complex task where small details may hold significant importance \cite{diffusionBeatGANs}.
Keeping in mind these limitations, we evaluated the quality of the generated CTPA volumes using four metrics:

\begin{enumerate}
    \item Fréchet Video Distance (FVD): FVD was proposed by Salimans et al. \cite{FVD}. It quantifies the dissimilarity between generated and real CTPA volumes by extracting image features using the I3D model \cite{I3Dmodel}, which is well suited to video datasets. This metric is only partially suited and holds inherent limitations as it is not fully suited to medical datasets.

\item PSNR: PSNR is often used to measure the quality of reconstructed digital signals \cite{PSNR}. The high dynamic range of the CTPA scan makes PSNR a good criterion for image quality evaluation.

\item SSIM: SSIM is a metric to measure the similarity of two images, including brightness, contrast and structure \cite{SSIM}. Compared to PSNR, SSIM better matches human’s subjective evaluation.

\item LPIPS \cite{LPIPS}: perceptual similarity (LPIPS) essentially computes the similarity between the activations of two image patches for some pre-defined neural network. This measure has been shown to match human perception well. A low LPIPS score indicates that image patches are perceptually similar.

\end{enumerate}

We computed these metrics by evaluating the generated samples against their corresponding ground truth samples in our test set. \textbf{Table \ref{tab:quantitative_results}} presents the quantitative generation results for the CTPA dataset. We recognize that these quantitative metrics alone may not fully capture the visual quality of our generated scans. Additional comparisons and analyses are provided in the subsequent Ablation Study section.

\begin{table}[ht!]
\caption{Quantitative evaluation of the synthesized test set samples using different metrics. FVD scores are multiplied by $10^3$.} \label{tab:quantitative_results}
\centering
\begin{tabular}{@{}cllll@{}}
\toprule
\multicolumn{1}{c}{Dataset} & \multicolumn{4}{c}{Metric} \\
\midrule \midrule
 & FVD $\downarrow$ & PSNR $\uparrow$ &  SSIM $\uparrow$ & LPIPS $\downarrow$\\
\midrule
CTPA & $0.29 \pm 0.22$ & $19.41 \pm 0.96$ & $0.58 \pm 0.08$ & $0.14 \pm 0.04$\\
\bottomrule
\end{tabular}
\end{table}

\paragraph*{Mapping the Latent Space of Generated Samples.}

\begin{figure}[hbt!]%
\centering
\begin{tabular}{c} 
\includegraphics[scale=0.4]{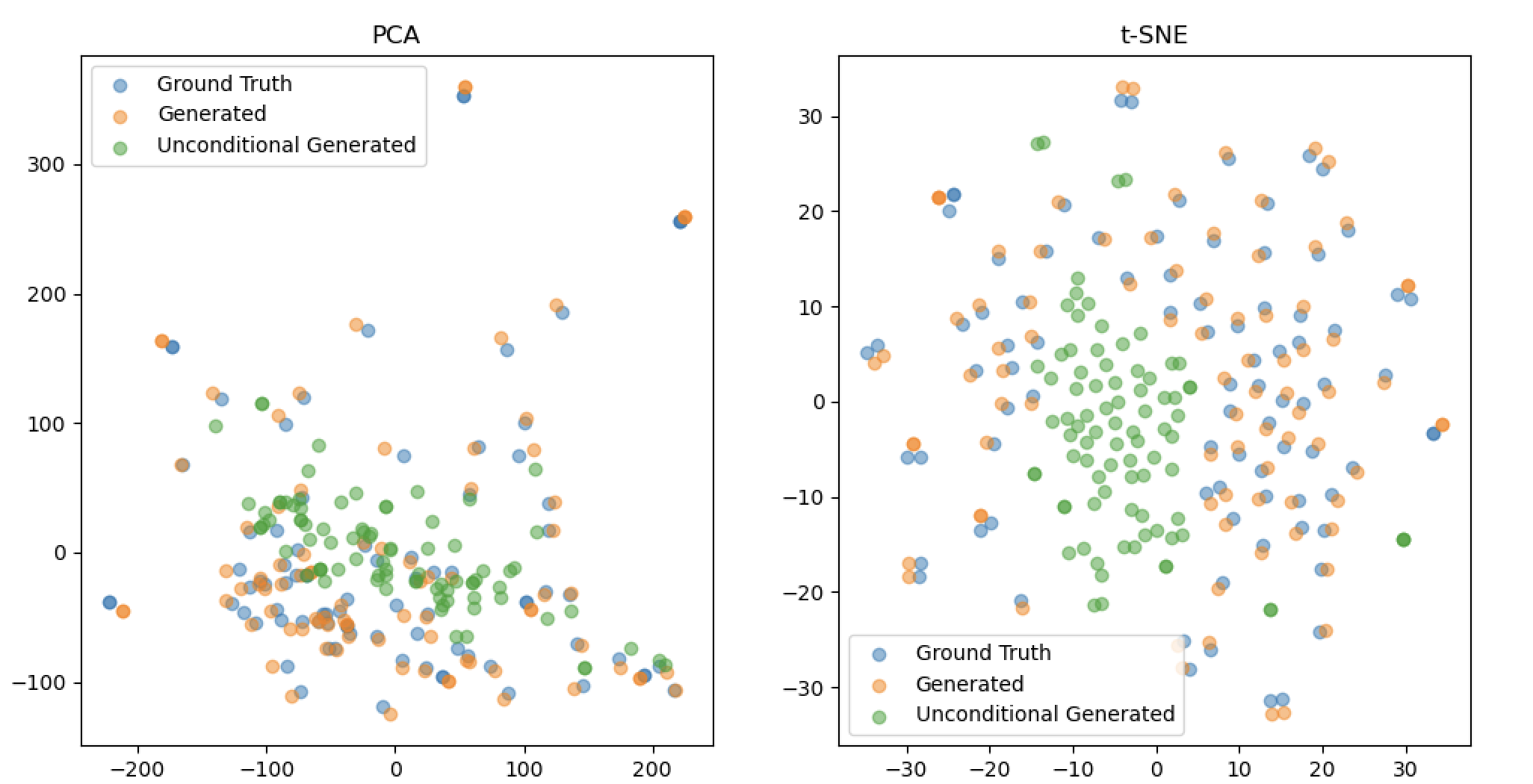} \\
\includegraphics[scale=0.4]{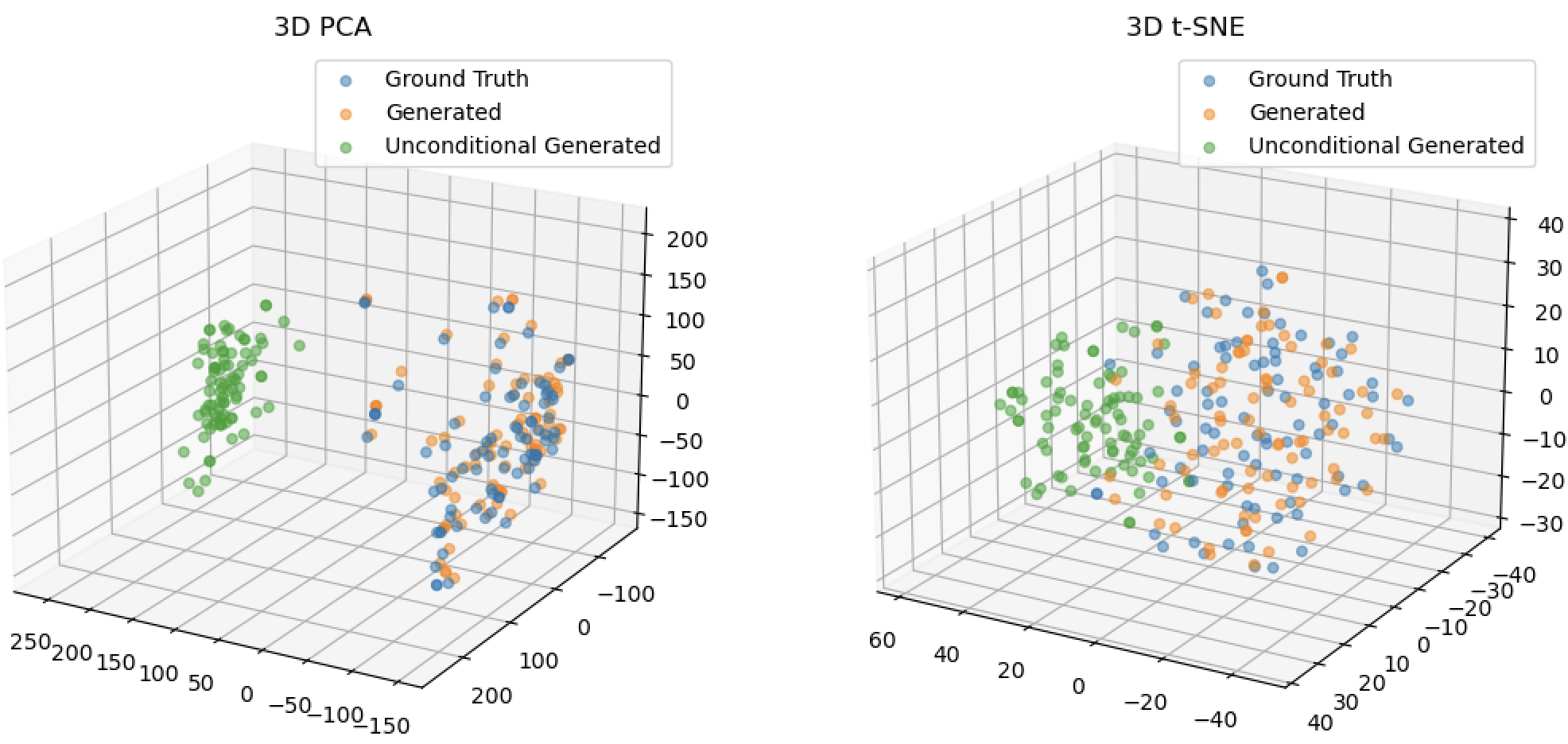} \\
\end{tabular}
\caption{t-SNE and PCA visualizations results of generated latents compared to the ground truth samples on the test set. The ground truth samples are marked in blue, the generated samples are marked in orange and the unconditional samples in green. 
}\label{fig:tsne_pca}
\end{figure}

In \textbf{Fig. \ref{fig:tsne_pca}} we present the visualization results of the t-SNE \cite{t-SNE} and PCA \cite{PCA} of the latents generated by the conditional model compared to their corresponding ground truth samples from the test set. For comparison, we also present the t-SNE and PCA of the latents generated from an unconditional model. As can be seen from the results, the generated samples closely follow the ground-truth samples. This is especially apparent in comparison to the unconditional samples. 

\paragraph*{Ablation study on model architecture choices.}

\begin{figure}[hbt!]%
\centering
\begin{tabular}{c} 
\includegraphics[scale=0.5]{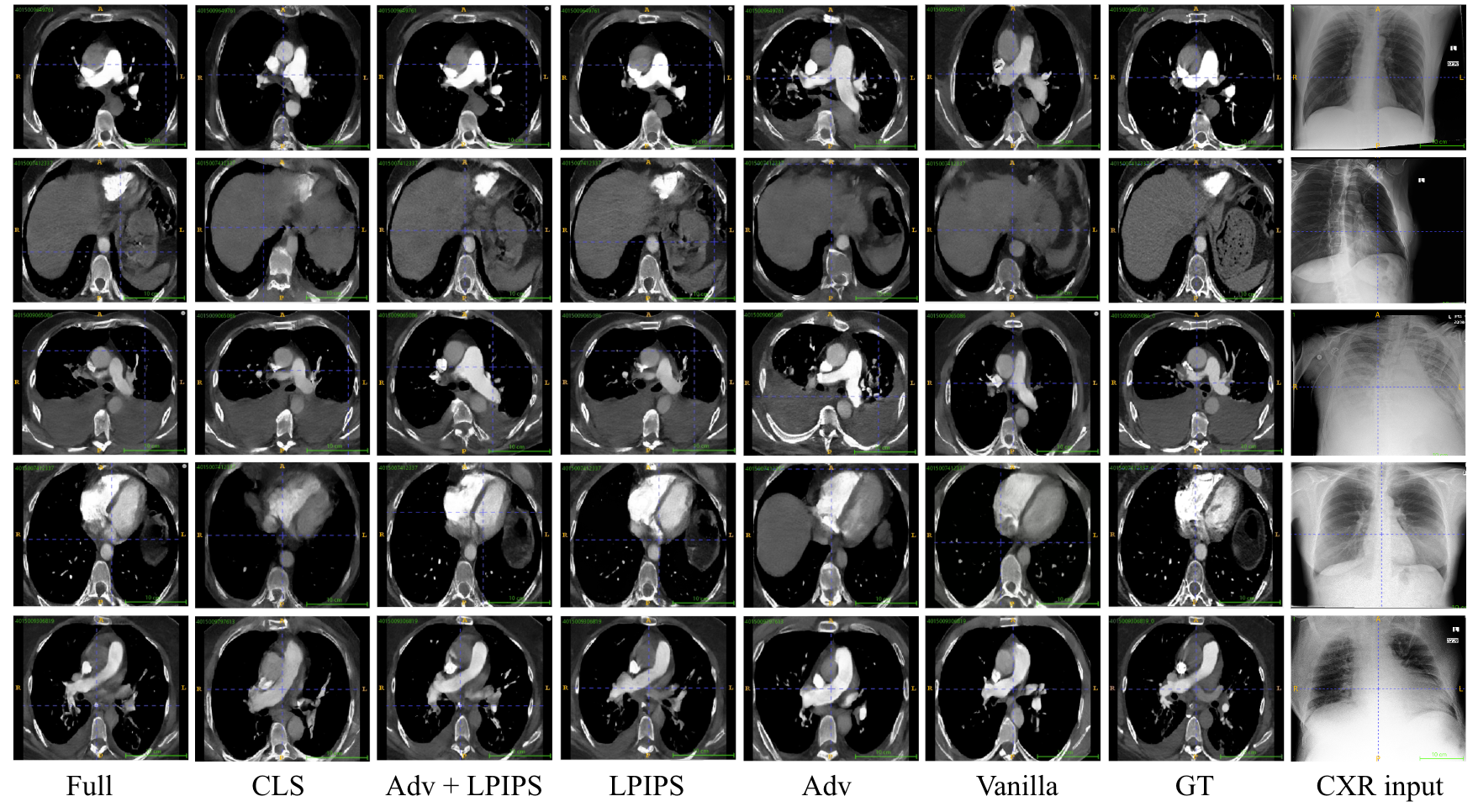} \\
\end{tabular}
\caption{Visual ablation of loss functions used for this study. Presented are different slices from generated 3D CTPA samples from our test set, which were trained using various loss functions and combinations. "Vanilla"= $l1$ loss, "Adv" = adversarial loss, "CLS" = classification loss and "Full"=weighted combination of all losses.
}\label{fig:ablation}
\end{figure}

Table \ref{tab:ablation_study} evaluates several design choices and the contribution of different network components. We examined different loss functions as alternatives for the model training. In addition to $L_{1}$ loss which minimizes the pixel distance between the generated sample and the ground truth, we compared the use of adding perceptual loss \cite{LPIPS}, adding adversarial loss using a discriminator and classification loss. As can be seen from the results, adding the perceptual loss improved the results compared to using only $L_{1}$ loss. 
However, while the addition of adversarial and classification loss did not enhance the quantitative metrics, they both played important roles for our model. Adding the adversarial loss  resulted in more visually realistic samples. The classification loss improved the classification results. This is presented in \textbf{Fig. \ref{fig:ablation}}. Additionally, we modified the pretraining process to include the RSPECT dataset not only in as an unconditional dataset for CTPA scans but also in a conditional setting by incorporating synthetic CXR images or DRRs created from the CTPA scans. Adding this pretraining step highly boosted our performance by all metrics.

We also present the results of experiments that failed to achieve optimal outcomes. We attempted to use classifier-free guidance (CFG) for both CXR conditioning and class label conditioning, but neither approach improved the results. Additionally, we experimented with variations in sample size. The spatial resolution of 512 likely yielded sub-optimal results due to our small sample size, and the resolution of 128 failed due to the limitations of the pretrained VAE we used, which was not trained on smaller image sizes. Lastly, we also attempted to use the pretrained 3D-VQGAN from medical diffusion \cite{medical_diffusion}, which was pretrained on the LIDC dataset, but it failed to improve results on the CTPA dataset.


\begin{table}[ht!]
\caption{Quantitative evaluation of the synthesized test set samples using different metrics for different model variants. FVD scores are multiplied by $10^3$.} \label{tab:ablation_study}
\centering
\begin{tabular}{@{}cllll@{}}
\toprule
Model Variant & FVD $\downarrow$ & PSNR $\uparrow$ &  SSIM $\uparrow$ & LPIPS $\downarrow$\\
\midrule \midrule
\multicolumn{5}{c}{Loss} \\
\midrule

L1 & $2.14 \pm 0.8$ & $11.61 \pm 0.91$ & $0.2 \pm 0.06$ & $0.48 \pm 0.07$ \\ 

L1 + LPIPS & $1.89 \pm 0.64$ & $14.78 \pm 0.60$ & $0.22 \pm 0.07$ & $0.34 \pm 0.06$ \\
L1 + ADV & $2.06 \pm 0.63$ & $11.53 \pm 0.78$ & $0.2 \pm 0.058$ & $0.45 \pm 0.052$\\ 
L1 + LPIPS + ADV & $1.94 \pm 0.63$ & $14.38 \pm 0.77$ &  $0.21 \pm 0.06$ & $0.36 \pm 0.048$\\
L1 + Classifier & $2.64 \pm 1.75$ & $12.03 \pm 0.82$ & $0.2 \pm 0.05$ &  $0.50 \pm 0.06$\\
\midrule
\multicolumn{5}{c}{Classifier Free Guidance} \\
\midrule
CFG & $2.16 \pm 1.78$ & $11.57 \pm 0.68$ & $0.19 \pm 0.04$ & $0.48 \pm 0.03$\\ 
label CFG & $2.05 \pm 0.78$ & $11.48 \pm 0.81$ & $0.19 \pm 0.06$ & $0.48 \pm 0.06$\\ 
\midrule
\multicolumn{5}{c}{Sample size} \\
\midrule
128 × 128 × 64 &  $2.88 \pm 1.2$ & $11.68 \pm 0.65$ & $0.22 \pm 0.09$ & $0.55 \pm 0.05$ \\
512 × 512 × 32 & $2.65 \pm 0.8$ & $11.88 \pm 0.77$ & $0.21 \pm 0.08$ & $0.55 \pm 0.05$ \\
\midrule
\multicolumn{5}{c}{Architecture components} \\
\midrule
3D-VQGAN & $4.01 \pm 1.4$ & $10.89 \pm 0.9$ & $0.12 \pm 0.07$ & $0.88 \pm 0.06$ \\
CLIP &  $3.74 \pm 1.2$ & $11.39 \pm 0.79$ & $0.14 \pm 0.08$ & $0.76 \pm 0.05$ \\
\midrule
\multicolumn{5}{c}{Full} \\
\midrule
X-ray2CTPA & $\textbf{0.29} \pm \textbf{0.22}$ & $\textbf{19.41} \pm \textbf{0.96}$ & $\textbf{0.58} \pm \textbf{0.08}$ & $\textbf{0.14} \pm \textbf{0.04}$\\

\bottomrule
\end{tabular}
\end{table}

In addition to the quantitative results, we also compare the visual differences of using different loss functions (\textbf{Fig. \ref{fig:ablation}}). All loss functions were trained as a weighted combination with the $l1$ loss. 
In general, it is evident that the generated samples are like the ground truth, although not identical.
Supporting the quantitative results, the LPIPS loss improves both the visual appearances of the generated samples and seem to guide the scans to be closer in structure and color contrast to the ground truth. The pretraining step using the DRRs as conditioning further enhances the LPIPS loss improvements and provides the most realistic images which are also the most close to the ground truth. The classification loss adds distortion to the samples, which is expected as its purpose is only to aid the classification task. The adversarial loss provides very realistic images but, in some cases, the generated sample is derived further away from the ground truth. The weighted combination of all losses (\textbf{Fig. \ref{fig:ablation}} - "Full") provides the best trade-off between all losses.

\paragraph*{Exploring the use of the 2D-VAE Vs. 3D-VQGAN.} 

We explore two alternatives as compression components for the latent diffusion process: a 3D-VQGAN applied to the entire volume and a 2D-VAE applied to individual 2D slices of the 3D volume, which are subsequently concatenated. Both compressors were used "off-the-shelf" without additional training on our dataset. The 3D-VQGAN was sourced from the medical diffusion model \cite{medical_diffusion} and trained on the LIDC dataset \cite{LIDC}. Table \ref{tab:quantitative_results_vae} presents the quantitative comparison between these two approaches. The results compare reconstructed samples from the test set, processed through each compression component, against the ground truth. The quantitative analysis shows that both methods perform well, with the 2D-VAE outperforming the 3D-VQGAN, despite not being trained on medical data. Notably, the FVD score of the 2D-VAE demonstrates that the continuity between slices within the 3D volume is preserved.

\begin{table}[ht!]
\caption{Quantitative evaluation of the reconstructed test set samples from the 2D-VAE.} \label{tab:quantitative_results_vae}
\centering
\begin{tabular}{@{}cllll@{}}
\toprule
\multicolumn{1}{c}{Compressing Component} & \multicolumn{4}{c}{Metric} \\
\midrule \midrule
 & FVD $\downarrow$ & PSNR $\uparrow$ &  SSIM $\uparrow$ & LPIPS $\downarrow$\\
\midrule
2D-VAE & $0.05 \pm 0.05$ & $31.00 \pm 2.5$ & $0.84 \pm 0.09$ & $0.03 \pm 0.02$\\
3D-VQGAN & $0.1 \pm 0.2$ & $26.35 \pm 2.8$ & $0.74 \pm 0.08$ & $0.05 \pm 0.03$\\
\bottomrule
\end{tabular}
\end{table}

\begin{figure}[hbt!]
\centering
\begin{tabular}{c}
\includegraphics[scale=0.55]{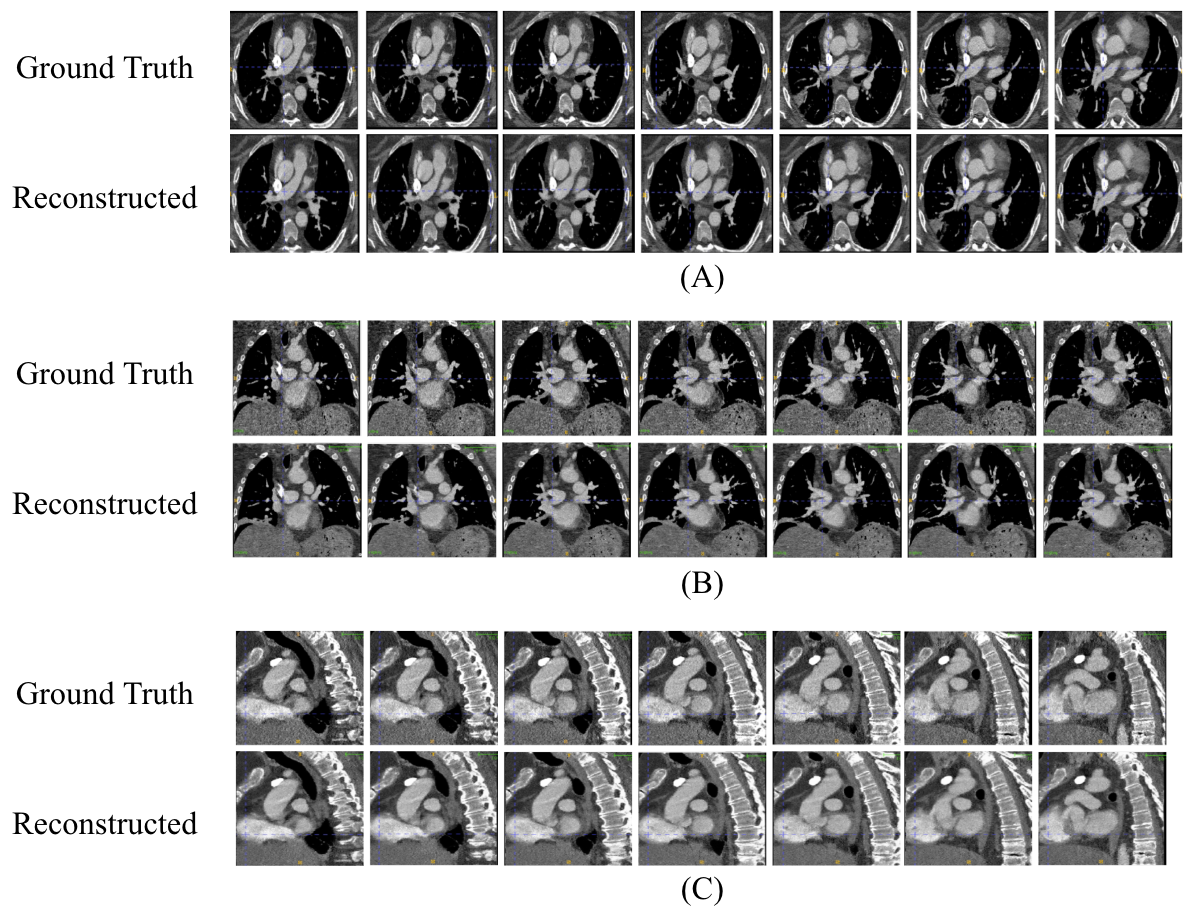}
\end{tabular}
\caption{Visual results of ground truth and reconstructed samples after applying 2D-VAE compression and decompression. Presented are consecutive slices from three views of the 3D volume ((A) axial, (B) coronal and (C) sagittal)}
\label{fig:qual_2dvae}
\end{figure}

\textbf{Fig. \ref{fig:qual_2dvae}} shows consecutive 2D slices from the axial, coronal, and sagittal views of the 3D volume, comparing the ground truth to the samples reconstructed by the 2D-VAE. The results indicate that the transitions between slices remain continuous across all views, and the inter-slice changes are preserved during the compression and decompression processes of the 2D-VAE.

\paragraph*{Generalizability.} 

We wanted to evaluate the ability of our model to generate other modalities in a similar setting. To do so, we fine-tuned our model on an additional dataset. We used chest CT studies from the publicly available Lung Image Database Consortium (LIDC) dataset \cite{LIDC}. This dataset contains 1010 individual studies (thoracic CT) . We noticed that in addition to the available CT scans, around 290 scans also included matching CXRs. We fine-tuned the model using low rank adaptation (LoRA) \cite{hu2021lora}. 

The quantitative generation results for the LIDC dataset are presented in \textbf{Table \ref{tab:quantitative_results_lidc}}. As can be seen, although trained on an extremely low number of samples - the evaluation results are high. 

\begin{table}[ht!]
\caption{Quantitative evaluation of the synthesized test set samples from LIDC dataset. FVD scores are multiplied by $10^3$.} \label{tab:quantitative_results_lidc}
\centering
\begin{tabular}{@{}cllll@{}}
\toprule
\multicolumn{1}{c}{Dataset} & \multicolumn{4}{c}{Metric} \\
\midrule \midrule
 & FVD $\downarrow$ & PSNR $\uparrow$ &  SSIM $\uparrow$ & LPIPS $\downarrow$\\
\midrule
LIDC & $0.97 \pm 0.71$ & $18.02 \pm 2.12$ & $0.53 \pm 0.06$ & $0.31 \pm 0.07$\\ 
\bottomrule
\end{tabular}
\end{table}

\begin{figure}[hbt!]
\begin{tabular}{c}  
\includegraphics[scale=0.50]{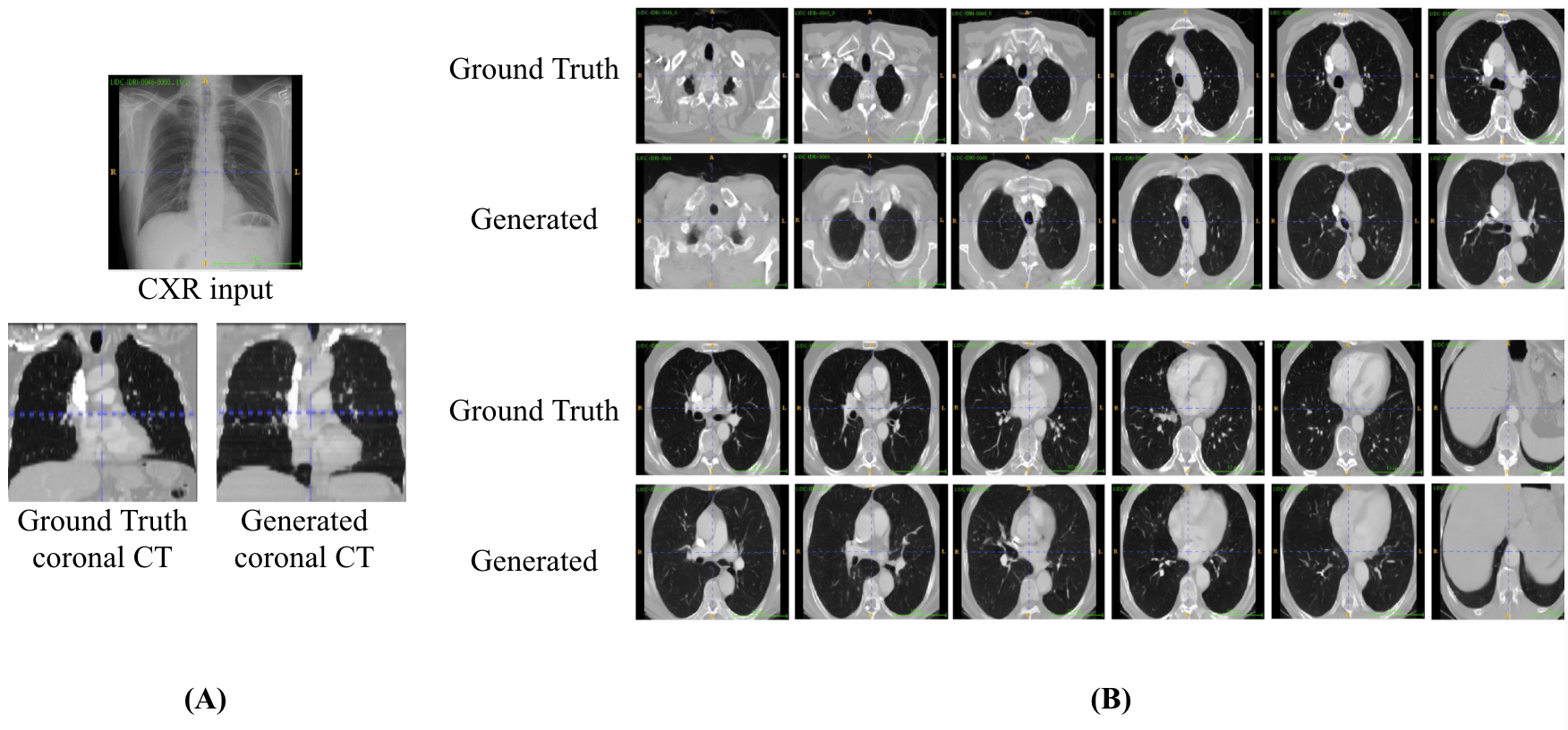} \\
\end{tabular}
\caption{Qualitative Results of CXR to 3D CT Generation with X-ray2CTPA model from LIDC dataset. (A) We present a generated 3D-CTPA sample from the test set, generated from a 2D CXR input. We compare slices from the generated 3D-CT scan to the ground truth sample. (B) The input CXR is presented along with the matching coronal slices from the matching CT and generated 3D-CT. This is the same sample as in the previous figure.}
\label{fig:sample_slices_lidc}
\end{figure}

In addition to the quantitative results we also provide a generation example of one scan compared to its corresponding ground truth. The results are presented in \textbf{Fig. \ref{fig:sample_slices_lidc}}.

\subsection*{User Studies}
We conducted two short expert evaluation studies, with the generated scans separately evaluated by two board certified radiologists.
\paragraph*{Generated Image Quality.}
In this study, experts were separately presented with 20 CTPA volumes, comprising an equal mix of 10 real and 10 synthetic ones. Readers were tasked with assessing the visual quality of these volumes and determining whether each volume was real or synthetic. Both readers accurately distinguished between most synthetic and real volumes, citing the inferior image quality of the synthetic scans compared to the real scans as the primary reason. As a followup, we asked the experts to rate the synthetic images in terms of anatomic correctness and slice consistency.  The readers were asked to rate 10 generated images from the test set. All readers rated the scans as mostly realistic with only minor unrealistic areas.  In addition, the images were considered consistent across slices with most images assessed as exhibiting only minor or no anatomic inconsistencies. We conclude from this initial study that our architecture can generate synthetic images that appear somewhat realistic to radiologists but further work needs to be done if we wish to achieve realistic-looking images. We hypothesize that better-quality output images can be generated once more data is used in the training phase. The small inconsistencies may be mitigated by adding a super-resolution stage, increasing the sample size, or using a pre-trained diffusion model and adapting it to our problem using tools such as ControlNet \cite{zhang2023adding}.
\paragraph*{PE categorization from CXR data - with and without Generated CTPA data.}
We compare radiologists' performance when interpreting conventional CXRs for PE categorization versus assessing PE from the generated CTPA images. In the initial phase with CXR alone (20 cases: 10 PE, 10 non-PE), the readers struggled to identify the confirmed emboli, producing an overall diagnostic accuracy close to chance levels (approximately 50\%), with very low sensitivity. This poor performance reflects the well-known limitation of chest radiography in diagnosing PE. In contrast, during the second phase, the same cases were reassessed, incorporating the corresponding generated 3D CTPA volumes. \textbf{Table \ref{tab:user_study}} summarizes the results of this study. Here, the radiologists correctly identified the majority of PE cases. The diagnostic accuracy improved substantially to 75\% (p-value = 0.025 and 0.033 using McNemar's test), with a clear increase in true-positive detections, demonstrating a significant boost over the CXR baseline. These results indicate that the synthesized CTPA imaging provided critical visual clues that were absent on CXRs, substantially enhancing the radiologists' ability to detect PE.

\begin{table}[ht!]
\caption{Expert evaluation of PE categorization task.} \label{tab:user_study}
\centering
\begin{tabular}{@{}cccc@{}}
\toprule
Experts & Accuracy  & Sensitivity & Specificity \\
\midrule
Reader 1 & $0.75$ & $0.8$  & $0.7$\\
Reader 2 & $0.75$ & $0.9$ & $0.6$  \\
\bottomrule
\end{tabular}
\end{table}



\subsection*{Ablation Study on CTPA Classifier}
We implemented two different classifiers for the real and synthetic CTPA scans. The first is the INSPECT classifier \cite{huang2023inspect}, which is implemented in the pixel space, and the second is a 3D DenseNet3D-121 classifier trained in the latent space of the diffusion model. The detailed architectural and training details for both classifiers are provided in the Methods section.

\textbf{Table \ref{tab:classification_results_ablation}} summarizes the comparison results of the two classifiers on the original and synthesized CTPA scans. The data set in its latent form is limited by the ability of the diffusion model to generate only a maximum of 64 frames along the z‑axis. Therefore, we benchmarked INSPECT with three sets of slices: 512, 128, and 64. Using all 512 slices, INSPECT achieves the strongest performance (AUC = 0.90; 95 \% CI [0.88–0.97]). As the slice count decreases, its AUC drops accordingly. The results of the latent classifier and the 64-slice INSPECT model receive similar results, where INSPECT slightly outperforms the latent classifier with a gain of 2\% 
On the synthesized scans, the latent-space classifier substantially surpasses the INSPECT classifier. We attribute this gap to the training strategy - during diffusion training, we use the relation between CXR and 3D CTPA to infuse the latent classifier with information unavailable to a model trained solely in pixel space.

\begin{table}[ht!]
\caption{Ablation on the CTPA classifier. We compare the average AUC scores (with a 95\%  confidence interval) of the PE classification from the original CTPAs and generated CTPAs of our test set.} \label{tab:classification_results_ablation}
\centering
\begin{tabular}{@{}llccc@{}}
\toprule
\multicolumn{1}{c}{Model} & \multicolumn{1}{c}{AUC} & Accuracy & Specificity & Sensitivity\\
\midrule \midrule
\multicolumn{5}{c}{Original CTPA data}\\
\midrule
CTPA INSPECT 512 slices & $0.90$  [$0.88$-$0.97$] $\pm 0.08$ & $84.05$ & $83.86$ & $82.74$ \\
CTPA INSPECT 128 slices & $0.88$ [$0.8$-$0.95$] $\pm 0.1$ & $79.5$ & $89.6$ & $73.91$ \\
CTPA INSPECT 64 slices & $0.85$ [$0.77$-$0.92$] $\pm 0.08$ & $82.3$ & $84.7$ & $74.9$\\
CTPA latent & $0.827$ [$0.75$-$0.92$] $\pm 0.05$ & $76.92$ & $78.18$ & $73.91$\\
\midrule
\multicolumn{5}{c}{Generated Data}\\
\midrule
Generated CTPA latent classifier & $0.80$ [$0.72$-$0.9$] $\pm 0.07$ & $74.0$ & $74.1$ & $73.91$ \\
Generated CTPA INSPECT classifier & $0.74$ [$0.68$-$0.82$] $\pm 0.06$ & $72.1$& $71.88$ & $71.9$\\
\bottomrule
\end{tabular}
\end{table}

\subsection*{PE Classifier Grad-CAM Visualization Results}

\begin{figure}[hbt!]%
\centering
\begin{tabular}{c} 
\includegraphics[scale=0.5]{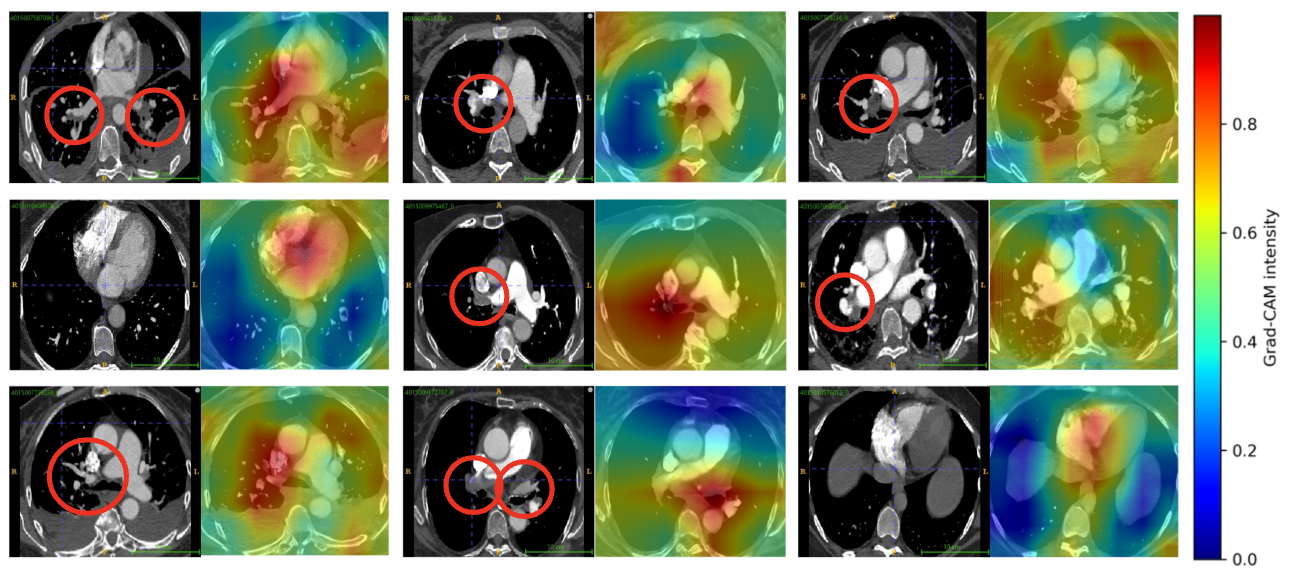}
\end{tabular}
\caption{\added{Interpretation of the model predictions using Grad-CAM visualization results of correctly classified examples. Presented are samples of Grad-CAM and original slices. For each sample, we paired the original CTPA slice (left) with the corresponding Grad-CAM overlay on the matching generated slice (right). Red circles indicate the precise PE locations as identified by an experienced radiologist.
High activations (red, yellow and green) indicate the focus areas that the model selected for classification. Note the high activations around PE clots and the heart area. 
}}\label{fig:grad_cam}
\end{figure}

We employed gradient back-propagation techniques for model interpretability. We highlighted the features selected by the model during prediction and identified locations and slices in the 3D-CTPA imaging scan that contributed the most to the classification using 3D gradient-weighted class activation mapping (Grad-CAMs) \cite{Grad_Cam}. For our implementation we used Captum open source package, an extensible library for model interpretability built on PyTorch \cite{captum}.
\textbf{Fig. \ref{fig:grad_cam}} illustrates the heatmap visualizations generated by back-propagating through the 3D PE classification model - highlighting areas that most influenced the model’s decision.
We anticipated that the model would focus on regions associated with PE as well as descriptors correlated with high-risk PE, such as the heart region and hepatic veins.
As expected, the network concentrated on areas containing PE clots and the heart (particularly the right and left ventricular (RV, LV) chambers). 
We note that the Grad-CAM predictions cover a broader region around the PE clots, resulting in less precise localization. We hypothesis this is due to performing the classification in the latent space, where spatial resolution is naturally lower.

\begin{figure}[hbt!]%
\centering
\begin{tabular}{c} 
\includegraphics[scale=0.5]{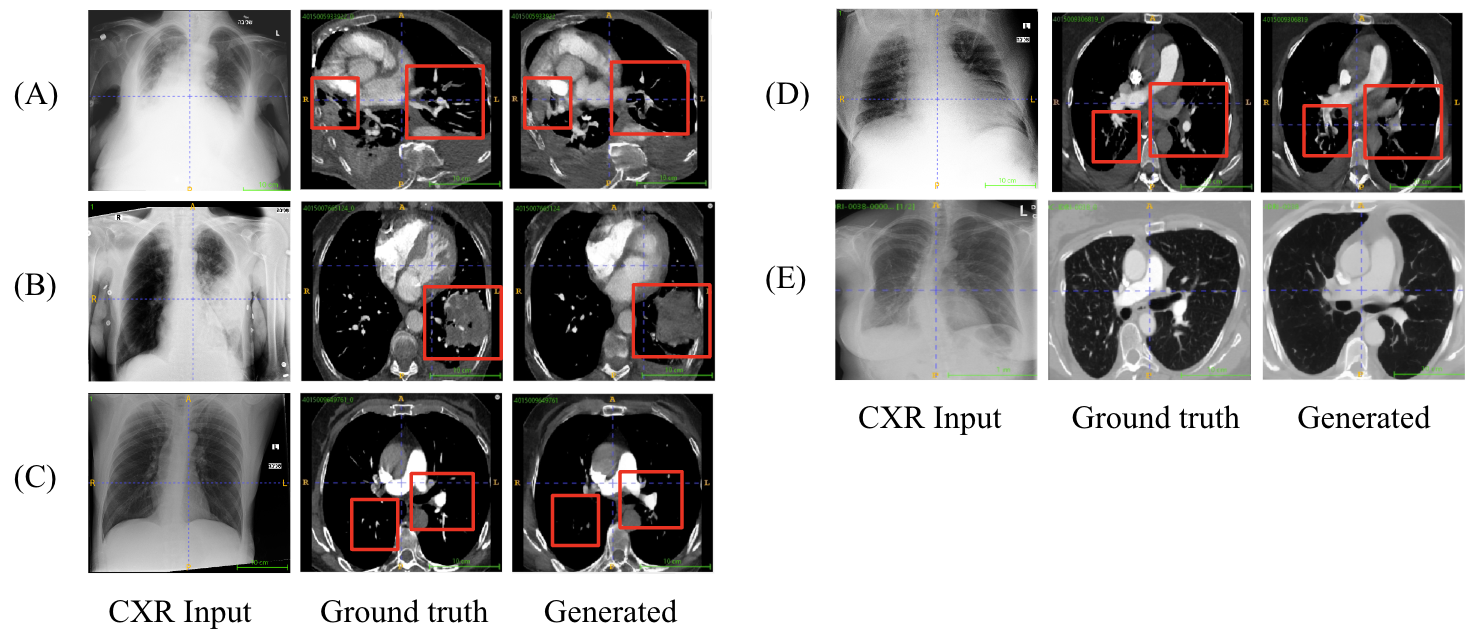} \\
\end{tabular}
\caption{Failed Examples and Hallucinations. Presented are example slices from our test set where we consider the model to fail. (A) Moderate pleural effusions and kyphoscoliosis pathologiesare apparent in both the CXR input and the ground truth sample. The generated sample shows the pleural effusions but they are not generated to be identical to the ground truth. (B) A lung cancer in the left lung is generated in the synthesized CTPA at the correct location and but is slightly different in texture and color. (C,D) Hallucinates the shape and locations of veins around the heart area. (E) Deformed right chest wall is absent. }\label{fig:hallucinations}
\end{figure}

\section*{Discussion}\label{sec:sec12}

The purpose of this work was to generate 3D CTPA scans from paired 2D CXRs in order to enhance PE classification from CXR images. To address this problem, we used a novel 3D diffusion model with BiomedCLIP Visual encoder, additional adversarial guidance, classifier and perceptual loss. We had many challenges when implementing this project: the number of samples in our dataset is relatively small, the size of each CTPA scan is large, the 3D nature of the CT scans enforces consistency throughout the generated 3D scan and between 2D slices which makes it hard to use the common 2D pretrained models. In addition, the 2D CXR data inputs are not registered or aligned to the CTPA scans. Finally, the application of PE is challenging due to the small size of the clots in the scans.

Our best performing model improved the CXR PE classification results by 11\%; Improving from an AUC of 0.691 [95\% CI: 0.54-0.77] to 0.803 [95\% CI: 0.72-0.9]. In addition, the specificity of the model improved, which in a clinical setting may potentially lead to a reduction of redundant CTPA scans performed for PE suspected patients. Unlike other diffusion studies in the medical field, this research uniquely applies generated data to address a real-life clinical challenge rather than focusing on the quality of the generated samples.
We were able to generate somewhat realistic, consistent and anatomically correct 3D scans. The generated 3D scans are not identical to the ground truth CTPAs but are similar and capture their structure. We experimented with the image size and additional modifications to the model and loss functions. 
As far as we know, this is the first model that attempts to convert 2D chest X-rays to CTPA scans. We also demonstrate the model's generalizability by fine-tuning it on an extremely small number of samples to achieve 2D CXR to 3D CT translation. 

The classification results presented not only substantiates the findings of this study but also highlights the value of the proposed methodology.
We present Grad-CAM visualizations that demonstrate our classifier focuses on anatomically and clinically relevant regions indicative of pulmonary embolism rather than on potential hallucinated features. Furthermore, our initial human observer study shows that the availability of generated 3D CTPA scans significantly enhances PE detection from CXR, underscoring the potential of these advanced imaging methods to improve diagnostic confidence when standard CXR imaging is insufficient.
We argue that the generation task acts as a valuable pre-text task for classification, serving as a form of regularization rather than the primary output. This perspective aligns with OpenAI's Image-GPT \cite{Image-GPT}, which demonstrates that generative image modeling is a promising route to learn high-quality unsupervised image representations. Furthermore, performing the classification within the 3D diffusion latent space reinforces this assertion.

We acknowledge the growing number of studies in the field of medical generative AI, particularly in Chest X-ray to CT translation. However, our work distinguishes itself in several key aspects: 1. To the best of our knowledge, this is the first study to address the challenge of converting CXRs to CT scans directly, rather than relying on DRRs. Unlike DRRs, which contain significantly more information and are inherently aligned with CT scans, CXRs present a greater challenge due to their limited information and lack of alignment. Our proposed method addresses this challenge without requiring extensive preprocessing or alignment between modalities, resulting in a more robust and adaptable approach. 2. Many generative models in this field are primarily focused on data generation and are rarely applied to solving specific downstream tasks, such as classification or segmentation. In contrast, this paper goes beyond data augmentation by directly demonstrating the utility of our approach in solving a classification problem, showcasing its practical application and impact.

We have tested multiple CXR-focused baselines, and were not able to compete with the generative-based results presented. On the training set these models reach a performance comparable to our pipeline; its shortfall appears once we test on unseen and the model’s ability to generalize. The key difference is geometric context: a single CXR provides only a 2D projection, whereas the diffusion step generates a 3D representation of the lungs. This volumetric view allows the downstream classifier to inspect clot-relevant patterns that are entirely collapsed in a projection image, leading to better generalization. Similar trends appear in other fields, in which introducing a generative step can provide the model with richer or more varied information than the raw inputs alone.Some examples are: ImageGPT \cite{Image-GPT}, shows that diffusion-style pretext tasks learn high-quality unsupervised latents that boost downstream accuracy; GAN-based synthetic augmentation in medical imaging routinely improves classification performance \cite{FRIDADAR2018321,düzyel2023dataaugmentationganincreases}. Likewise, in NLP, generative pre-training of language models such as BERT and GPT substantially lifts text-classification benchmarks \cite{ding2024data}. These generative steps expand the discriminative feature space, enhance generalization, and ultimately yield representations that surpass direct classification on raw images.

This study is subject to a few significant limitations. It was designed as a single-center, retrospective study, which is associated with well-known shortcomings and inherent limitations. In addition, generalizing the methods and results of our study to other predictive tasks with different modalities should be examined carefully, as our comparison and analysis were limited to CXR and CTPA modalities and to the task of PE classification. Finally, while our model generates 3D CTPA scans which are close to the ground truth, they are by no means identical to them and currently cannot be used to replace CTPA scans. 

While X-ray2CTPA model can generate high-quality 3D CTPA samples that closely resemble the ground truth in structure and appearance, there are instances where it slightly diverges or fails to accurately replicate the ground truth. Some examples from our test set are presented in \textbf{Fig. \ref{fig:hallucinations}}. The main areas where the model fails are around small veins or when rare abnormalities or additional pathologies (other than PE) are presented. In some of these cases the model generates the pathology in some variant but may replicate it in a wrong location or fail to generate the pathology whatsoever. We hypothesis that this issue may be caused by the high variability nature of the CTPA samples and due to the fact, that in many cases PE is secondary to other patient comorbidities. We believe that training the model with a larger dataset may solve this issue in future work.
We hypothesize that in order for the generated scans to be used as actual replacements for the original CTPAs, we need a much larger dataset.

\begin{figure}[hbt!]%
\centering
\begin{tabular}{c} 
\includegraphics[scale=0.3]{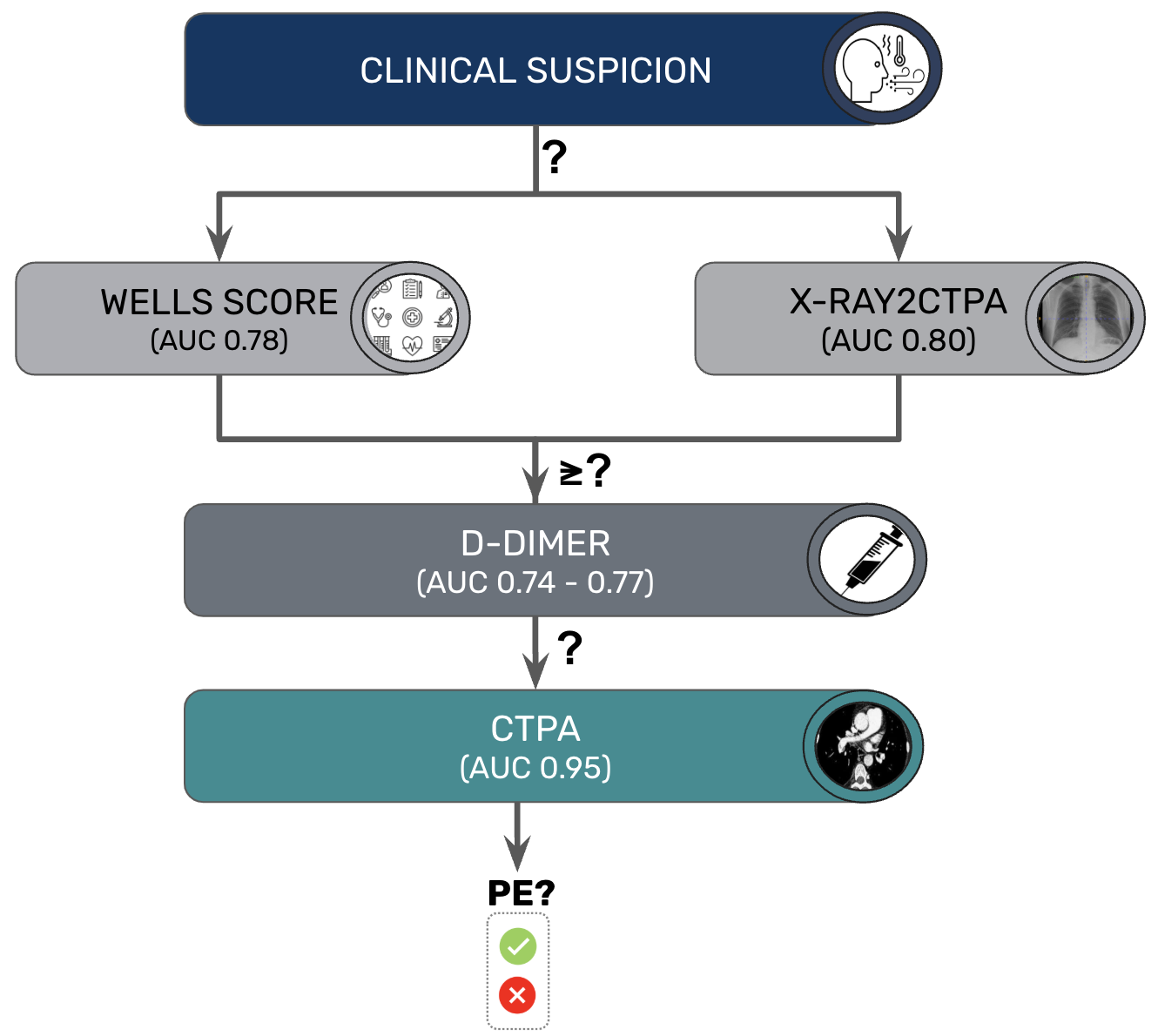} \\
\end{tabular}
\caption{A possible clinical workflow in which X-ray2CTPA can be used. The X-ray2CTPA model can be used along with the Wells Score for decision whether to perform D-Dimer or not. In this setup the model serves as a practical triage support tool (Icon images from Microsoft PowerPoint 2023.).}\label{fig:proposed_pipeline}
\end{figure}

We acknowledge that, at its current performance level, our model does not meet the stringent sensitivity standards typically required for clinical triage. Rather, our work represents an initial proof of concept demonstrating the potential utility of generated 3D CTPA scans in guiding preliminary decisions. For context, the widely used Wells Score achieves an AUC of approximately 0.78 \cite{wells}, so our X-ray2CTPA approach could serve as an adjunct - augmenting the Wells Score rather than replacing CT imaging - to refine pre-test probability and guide the decision to perform a D-dimer assay from a simple chest X-ray. As the technology matures, incorporating larger datasets, improved architectures, and refined training strategies may increase both sensitivity and specificity. This incremental progress could eventually enable the model to serve as a practical triage support tool, flagging likely PE cases for more urgent follow-up imaging while reducing unnecessary scans in low-risk scenarios \textbf{Fig. \ref{fig:proposed_pipeline}} illustrates a proposed workflow combining Wells Score, generated CTPA, and D dimer testing.

For future work, we intend to apply our model to alternative paired modalities to test the model's ability to generalize to other domains. X-ray2CTPA demonstrates promising results in generating realistic and accurate 3D CTPA scans, paving the way for a future where AI-powered medical imaging can transform healthcare by making advanced diagnostic tools more accessible and affordable for patients worldwide.

\section*{Methods}

\subsection*{X-ray2CTPA diffusion model}
\begin{figure}[hbt!]%
\centering
\begin{tabular}{c} 
\includegraphics[scale=0.65]{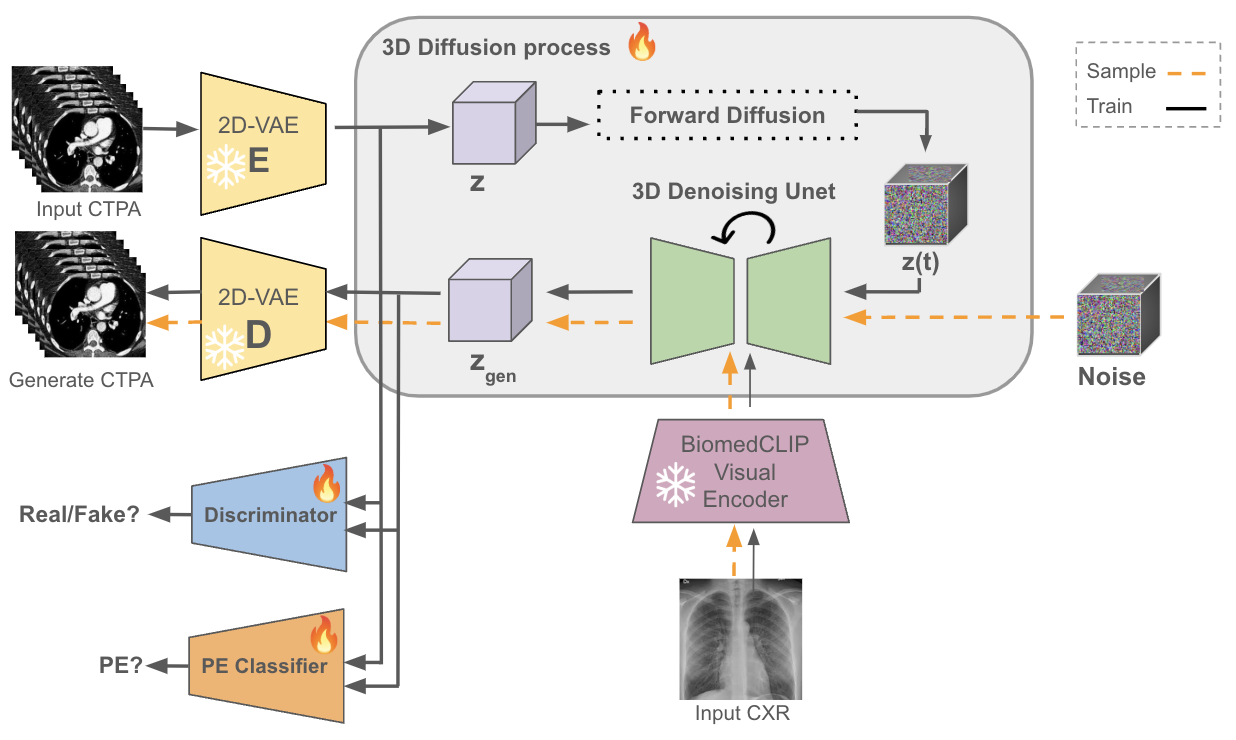} \\
\end{tabular}
\caption{X-ray2CTPA model architecture: Black arrows - Training: The CTPA scan are fed slice by slice to the 2D-VAE in a sequential manner and compressed by a factor of 8. The compressed slices are concatenated. The latent representation is noised in the forward diffusion process and is then used as input to the 3D de-noising U-net model. The CXR acts as conditioning to the model, fed into the CLIP Vision encoder. Bottom - Sampling: A noise latent and CXR are the inputs to the model and the output is the reconstructed 3D CTPA scan.
}\label{fig:model}
\end{figure}

The limited number of samples in our dataset, their substantial size and 3D nature, were the main considerations for our model architecture selection. The model is constructed from a two-step approach: first, the images are encoded into a low-dimensional latent space using a frozen pretrained 2D-VAE and subsequently, a 3D diffusion probabilistic model is trained on the latent representation of the data. The compressed nature of the low-dimensional latent space inherently enables faster training and inference and alleviates the problem of limited computational resources. An outline of our model architecture is presented in \textbf{Fig. \ref{fig:model}}.

The 3D-CTPA scans were fed to the 2D-VAE sequentially slice by slice and then concatenated before being fed into the 3D diffusion model. For the conditional setting, we added 2D CXR scans as conditioning to the diffusion model. We hypothesize that the CXR to CTPA modality shift is more like text to image conversion rather than image-to-image in-painting. Therefore, we applied the conditioning by concatenation of the CXR embedding throughout the 3D U-net layers, rather than applying it as concatenation to the noise as input of the U-net. The CXR was encoded to a 512-vector using BiomedCLIP Visual encoder \cite{medclip}. We further improve the 3D scan generation by imposing a 4 layered 3D discriminator at the output of the diffusion process. In addition, a DenseNet3D-121 \cite{huang2018densely} classification was also applied on the generated latents to introduce label information and improve classification results. 

\subsection*{PE Classification.}
We employed two single-modality encoders for PE classification. One encoder processes CTPA data, while the other handles CXR data. These encoders serve as baseline classification models. In addition, we leverage the trained CTPA encoder as a pretrained model to classify PE on the CTPA scans generated by our X-ray2CTPA model.

\paragraph*{CTPA Classifier.}
The trained model is introduced during the training phase of the X‑ray2CTPA model as a classification loss, and it is subsequently employed to evaluate the generated CTPA scans once training is complete. In addition, the classifier is applied to the generated latents during training to inject label information and enhance CXR classification results.

We encountered several challenges that guided our selection of the CTPA classifier architecture. Our dataset is relatively small and provides only a single binary label for each entire 3D volume. We were further constrained by the diffusion model’s ability to generate only a maximum of 64 frames along the z‑axis. We explore the optimal CTPA classifier for both real and generated scans. We applied two complementary approaches, one operating directly in pixel space and the other in the diffusion model’s latent space: 1. \textit{INSPECT pixel-space classifier \cite{huang2023inspect}}: Although INSPECT’s published performance is not state-of-the-art (SOTA) for PE classification, it embodies the current best-practice strategy for this task. The model is pre-trained with 2D label supervision for every slice in a 3D volume, and can process an unlimited number of slices without resampling. The model is comprised of a two-phase pipeline: At first, a 2D ResNet-101 is trained using 2D labels from each slice of the 3D scan. This trained model is then frozen and used as a feature extractor. In the second phase, the extracted features are fed into a sequential LSTM model for a final prediction. We note that we were unable to fine-tune the ResNet backbone on our dataset, as our labels are provided only at the 3D scan level. As a result, we could only train the LSTM part of the model.
2. \textit{Latent space classifier}: 
We also trained a 3D DenseNet-121 on the latent representations produced by the 2D VAE that is used in our diffusion model. This network operates on the compressed (scaled-down) version of each CTPA volume, enabling efficient learning while preserving volumetric context.

For the latent classifier, in the first stage, the classifier was trained on the original CTPA dataset (excluding our held‑out test set) using a scaled‑down latent representation derived through the same processing pipeline as the generated latents from the X‑ray2CTPA diffusion model. Once the classifier was trained, we incorporated it into the X‑ray2CTPA model as a classification loss. During inference, we generated CTPA latents for the entire dataset and fine‑tuned the same model for evaluation.

\paragraph*{CXR Classifier.}

We evaluated several pre-trained models for the CXR encoder and fine-tuned them for the specific task of PE classification: 1. \textit{RadImageNet} \cite{radimagenet}, a ResNet-50 based model pretrained on millions of radiologic images. 2. \textit{BiomedCLIP Vision encoder} \cite{medclip}. The BiomedCLIP visual encoder, part of the BiomedCLIP foundation model, was pretrained on a biomedical dataset of 15 million figure-caption pairs from PubMed Central. It achieves SOTA performance across a variety of medical vision-language processing (VLP) tasks, including cross-modal retrieval, image classification, and visual question answering. This is the same encoder that was used in the X-ray2CTPA model for CXR conditioning. 3. \textit{DINOv2} \cite{dinov2}, one of the most powerful existing 2D foundation model. DINOv2 is a self-supervised distillation method applied on Vision Transformers (ViTs) ~\cite{transformer}. The model's self-supervised ViT features contain explicit information about the semantic segmentation of an image and the extracted features can be used for many downstream tasks.
The best performing encoder was the RadImageNet model, therefore we chose it as our baseline.

\subsection*{Training Strategy}

The diffusion model was pretrained on the RSPECT dataset \cite{RSPECT}. Then, the diffusion model was fine-tuned on our dataset in a two-stage setting. First, the model was fine-tuned in an unconditional manner, without any CXR conditioning. In the second stage, the CXR conditioning was added, and the model was further trained. Even though the datasets were relatively small, we found that each model converged and generated realistic synthetic images without a lot of hyperparameter finetuning.
During training, the 2D-VAE and BiomedCLIP visual encoder remained frozen, while the 3D diffusion model, classifier, and discriminator were fine-tuned and trained further.

We utilized a pretrained 2D-VAE, fine-tuned from the original KL-f8 autoencoder, which demonstrated improved face reconstruction capabilities \cite{2dvae1,2dvae2}. Given its exceptional feature extraction capabilities on our dataset, the 2D-VAE required no additional fine-tuning and remained frozen throughout the training process. 
The BiomedCLIP visual encoder, part of the BiomedCLIP foundation model, was pretrained on a biomedical dataset of 15 million figure-caption pairs from PubMed Central. We selected BiomedCLIP for its SOTA performance across a variety of medical vision-language processing (VLP) tasks, including cross-modal retrieval, image classification, and visual question answering.
For the discriminator, we employed the medical-pretrained version of MONAI Generative \cite{monai-gen}, using its Patch-GAN discriminator architecture \cite{patch-disc}, which was further fine-tuned during training.

To address the challenge of limited sample size, we pretrained the diffusion model on the RSPECT dataset - the largest publicly available annotated PE dataset. RSPECT, from the 2020 RSNA Pulmonary Embolism Detection Kaggle Challenge, comprises over 7,000 3D CTPA studies from five international research centers \cite{RSPECT}. 
The classifier used in the X-ray2CTPA diffusion model pipeline was performed in the latent space and pretrained on latents extracted from the RSPECT dataset using the 2D-VAE. This classifier was trained separately before integration into the full X-ray2CTPA diffusion model setting.

\subsection*{Dataset}
Our cohort was constructed from 898 patients who underwent both a chest X-ray scan and a chest CT scan with contrast agent - CTPA within a 24-hour difference and from a single center. The chest CT volumes were annotated by board certified radiologists for a single label determining whether the patients had PE or not. The CTPA scans were anonymized and categorized by expert opinion indicating whether PE existed or not (no severity indication and no marked PE segmentation) according to the radiologist’s report. During the anonymization process, each patient is assigned a unique ID that links their various tests and prevents data leakage.
For the evaluation of the models, we used a stratified sixfold cross-validation. 305 of the samples (34 \%) were marked positive for PE. The dataset specifications are presented in Table \ref{tab:tab_dataset}. 

The institutional review board (IRB) of Sheba Medical Center approval was granted for this work. The need for informed consent was waived by the ethics committee. All methods and experiments were performed in accordance with relevant guidelines and regulations. The research was performed in accordance with the Declaration of Helsinki.

\begin{table}[ht!]
\caption{Dataset specifications}\label{tab:tab_dataset}
\centering
\begin{tabular}{@{}ll@{}}
\toprule
\multicolumn{1}{c}{Category} & \multicolumn{1}{c}{Numbers (\%))} \\ 
\midrule \midrule
Total & $898$ ($100$ \%) \\
\hline
Train &  $724$ ($80.6$ \%) \\
Validation & $88$ ($9.8$ \%) \\
Test & $86$ ($9.6$ \%) \\
\hline
Patients positive for PE & $305$ ($34$ \%) \\
Patients negative for PE & $593$ ($66$\%) \\
\bottomrule
\end{tabular}
\end{table}

\subsection*{Data pre-processing} 
The 3D CTPA scans had a spacial resolution of 512 × 512 with varying number of axial slices. All scans were resampled to a standard uniform voxel spacing  of 1 mm in all dimensions. The pixel values were converted to Hounsfield units (HU) and windowed to the range of $(-100 HU, +900 HU)$, to enhance the lower and upper boundaries of the area of interest limits of the HU scale. Finally, the scans were cropped to the region of the lung area using lung segmentation and resized to a size of 256 × 256 × 128 voxels (height × width × depth). Due to this preprocessing step, the axial slices of the scans seem a bit deformed. We min–max normalized the scans to the range of ($-1$,$1$). To achieve a reduced length without additional spacial manipulation of the data, the scans were divided into two segments of 256 × 256 × 64 each by selecting every alternate slice. Additionally, we augmented all datasets by vertically flipping the images during training with a probability of $50\%$.

The 2D CXR scans were normalized and resized to the size of 224 × 224 × 3 to match the size of the BiomedCLIP Vision encoder \cite{medclip}.
Although the 2D CXR and 3D CTPA scans are paired, we did not use any pre-registration procedures in our pipeline.

\subsection*{Loss functions}

We utilize multiple loss functions to improve its training, each with a distinct purpose. The overall loss is a combination of $l_1$, LPIPS \cite{LPIPS} weighted by $\lambda_1$, a discriminator loss, weighted by $\lambda_2$ and a classification loss weighted by $\lambda_3$. $l1$ loss ensures the generated images closely match the ground truth. LPIPS, leveraging the 2D ResNet-50 based RadImageNet architecture \cite{radimagenet}, enhances the model's grasp of complex semantic details and ensures perceptual similarity. 
The discriminator loss further ensures the generated images' realism, and the classifier loss guides the samples to the PE classification task. $\lambda_{1}, \lambda_{2}, \lambda_{3}$ were set to $0.01, 0.001$ and $0.08$ respectively. The loss can be summarized as follows:
\begin{equation}
loss = l_1 + \lambda_{1} * l_{LPIPS} +  \lambda_{2} * l_{adv} +  \lambda_{3} * l_{cls}
\label{eq:myEquation}
\end{equation}

\subsection*{Statistical Analysis}

The performance evaluation of the test set covered AUC, sensitivity (also known as recall), specificity, accuracy, the positive predictive value (PPV, also known as precision), and the negative predictive value (NPV). The DeLong method \cite{delong} was used to calculate the 95\% confidence intervals for the AUC. The probability thresholds for predicting positive samples were determined by the sensitivities and specificities of the validation set, which ensured high specificities while keeping reasonable sensitivities. These thresholds were determined by Youden's index \cite{youden}, which finds the optimal joint sensitivity and specificity.

\subsection*{Implementation Details} 
All models were trained on a single NVIDIA RTX A5000 GPU, and each took about a week of training. The code for our model was based on Medical Diffusion \cite{medical_diffusion} with various alterations. The training parameters included a base learning rate of $1e^{-4}$ for the unconditional setting and when pretraining using RSPECT dataset and a learning rate of $1e^{-5}$ when fine-tuned for the final conditional model. An additional pretraining step was added in which we used the RSPECT dataset for pretraining but in a conditional setting. We used artificial CXR images or DRRs as conditioning. The DRRs were created using DiffDRR \cite{DiffDRR} Python package for differentiable X-ray rendering. All were trained with a batch size of 2 with AdamW optimizer and Cosine Anneal scheduler. We used DDPM sampling with 1000 steps.


\begin{thebibliography}{10}
\urlstyle{rm}
\expandafter\ifx\csname url\endcsname\relax
  \def\url#1{\texttt{#1}}\fi
\expandafter\ifx\csname urlprefix\endcsname\relax\def\urlprefix{URL }\fi
\expandafter\ifx\csname doiprefix\endcsname\relax\def\doiprefix{DOI: }\fi
\providecommand{\bibinfo}[2]{#2}
\providecommand{\eprint}[2][]{\url{#2}}

\bibitem{xray}
\bibinfo{author}{Pfeiffer, D.}, \bibinfo{author}{Pfeiffer, F.} \& \bibinfo{author}{Rummeny, E.}
\newblock \bibinfo{journal}{\bibinfo{title}{Advanced x-ray imaging technology.}}
\newblock {\emph{\JournalTitle{Recent Results Cancer Res}}} \textbf{\bibinfo{volume}{216}}, \bibinfo{pages}{3--30} (\bibinfo{year}{2020}).

\bibitem{ct}
\bibinfo{author}{Ferrara, R.} \& \bibinfo{author}{Mansi, L.}
\newblock \bibinfo{journal}{\bibinfo{title}{Paul suetens (ed): Fundamentals of medical imaging (2nd edition)}}.
\newblock {\emph{\JournalTitle{European Journal of Nuclear Medicine and Molecular Imaging}}} \textbf{\bibinfo{volume}{38}}, \bibinfo{pages}{409--409} (\bibinfo{year}{2011}).

\bibitem{ct2}
\bibinfo{author}{Lo, P.} \emph{et~al.}
\newblock \bibinfo{journal}{\bibinfo{title}{Extraction of airways from ct (exact'09).}}
\newblock {\emph{\JournalTitle{IEEE Trans Med Imaging}}} \textbf{\bibinfo{volume}{31}}, \bibinfo{pages}{2093--2107} (\bibinfo{year}{2012}).

\bibitem{positive_pe}
\bibinfo{author}{Righini, M.}, \bibinfo{author}{Robert-Ebadi, H.} \& \bibinfo{author}{Le~Gal, G.}
\newblock \bibinfo{journal}{\bibinfo{title}{Diagnosis of acute pulmonary embolism.}}
\newblock {\emph{\JournalTitle{J Thromb Haemost}}} \textbf{\bibinfo{volume}{15}}, \bibinfo{pages}{1251--1261} (\bibinfo{year}{2017}).

\bibitem{hendy2023good}
\bibinfo{author}{Hendy, A.} \emph{et~al.}
\newblock \bibinfo{journal}{\bibinfo{title}{How good are gpt models at machine translation? a comprehensive evaluation}}.
\newblock {\emph{\JournalTitle{arXiv preprint arXiv:2302.09210}}}  (\bibinfo{year}{2023}).

\bibitem{smith}
\bibinfo{author}{Ramdurai, B.}
\newblock \bibinfo{title}{The impact, advancements and applications of generative ai} (\bibinfo{year}{2023}).

\bibitem{ai_in_medical}
\bibinfo{author}{Prevedello, L.~M.} \emph{et~al.}
\newblock \bibinfo{journal}{\bibinfo{title}{Challenges related to artificial intelligence research in medical imaging and the importance of image analysis competitions.}}
\newblock {\emph{\JournalTitle{Radiol Artif Intell}}} \textbf{\bibinfo{volume}{1}}, \bibinfo{pages}{e180031} (\bibinfo{year}{2019}).

\bibitem{x2ct}
\bibinfo{author}{Ying, X.} \emph{et~al.}
\newblock \bibinfo{title}{X2ct-gan: reconstructing ct from biplanar x-rays with generative adversarial networks}.
\newblock In \emph{\bibinfo{booktitle}{Proceedings of the IEEE/CVF conference on computer vision and pattern recognition}}, \bibinfo{pages}{10619--10628} (\bibinfo{year}{2019}).

\bibitem{mednerf}
\bibinfo{author}{Corona-Figueroa, A.} \emph{et~al.}
\newblock \bibinfo{title}{Mednerf: Medical neural radiance fields for reconstructing 3d-aware ct-projections from a single x-ray}.
\newblock In \emph{\bibinfo{booktitle}{2022 44th Annual International Conference of the IEEE Engineering in Medicine \& Biology Society (EMBC)}}, \bibinfo{pages}{3843--3848} (\bibinfo{organization}{IEEE}, \bibinfo{year}{2022}).

\bibitem{kazerouni2022diffusion}
\bibinfo{author}{Kazerouni, A.} \emph{et~al.}
\newblock \bibinfo{journal}{\bibinfo{title}{Diffusion models for medical image analysis: A comprehensive survey}}.
\newblock {\emph{\JournalTitle{arXiv preprint arXiv:2211.07804}}}  (\bibinfo{year}{2022}).

\bibitem{medical_diffusion}
\bibinfo{author}{Khader, F.} \emph{et~al.}
\newblock \bibinfo{title}{Medical diffusion - denoising diffusion probabilistic models for 3d medical image generation}, \doiprefix\url{10.48550/ARXIV.2211.03364} (\bibinfo{year}{2022}).

\bibitem{dispr}
\bibinfo{author}{Waibel, D.~J.}, \bibinfo{author}{R{\"o}ell, E.}, \bibinfo{author}{Rieck, B.}, \bibinfo{author}{Giryes, R.} \& \bibinfo{author}{Marr, C.}
\newblock \bibinfo{title}{A diffusion model predicts 3d shapes from 2d microscopy images}.
\newblock In \emph{\bibinfo{booktitle}{2023 IEEE 20th International Symposium on Biomedical Imaging (ISBI)}}, \bibinfo{pages}{1--5} (\bibinfo{organization}{IEEE}, \bibinfo{year}{2023}).

\bibitem{med-ddpm}
\bibinfo{author}{Dorjsembe, Z.}, \bibinfo{author}{Pao, H.-K.}, \bibinfo{author}{Odonchimed, S.} \& \bibinfo{author}{Xiao, F.}
\newblock \bibinfo{journal}{\bibinfo{title}{Conditional diffusion models for semantic 3d medical image synthesis}}.
\newblock {\emph{\JournalTitle{arXiv preprint arXiv:2305.18453}}}  (\bibinfo{year}{2023}).

\bibitem{makeavolume}
\bibinfo{author}{Zhu, L.} \emph{et~al.}
\newblock \bibinfo{title}{Make-a-volume: Leveraging latent diffusion models for cross-modality 3d brain mri synthesis}.
\newblock In \emph{\bibinfo{booktitle}{International Conference on Medical Image Computing and Computer-Assisted Intervention}}, \bibinfo{pages}{592--601} (\bibinfo{organization}{Springer}, \bibinfo{year}{2023}).

\bibitem{generatect}
\bibinfo{author}{Hamamci, I.~E.} \emph{et~al.}
\newblock \bibinfo{journal}{\bibinfo{title}{Generatect: Text-guided 3d chest ct generation}}.
\newblock {\emph{\JournalTitle{arXiv preprint arXiv:2305.16037}}}  (\bibinfo{year}{2023}).

\bibitem{paulson2024xprospect}
\bibinfo{author}{Paulson, B.} \emph{et~al.}
\newblock \bibinfo{journal}{\bibinfo{title}{Xprospect: Ct volume generation from paired x-rays}}.
\newblock {\emph{\JournalTitle{arXiv preprint arXiv:2403.00771}}}  (\bibinfo{year}{2024}).

\bibitem{x_diffusion}
\bibinfo{author}{Bourigault, E.}, \bibinfo{author}{Hamdi, A.} \& \bibinfo{author}{Jamaludin, A.}
\newblock \bibinfo{journal}{\bibinfo{title}{X-diffusion: Generating detailed 3d mri volumes from a single image using cross-sectional diffusion models}}.
\newblock {\emph{\JournalTitle{arXiv preprint arXiv:2404.19604}}}  (\bibinfo{year}{2024}).

\bibitem{delong}
\bibinfo{author}{DeLong, E.~R.}, \bibinfo{author}{DeLong, D.~M.} \& \bibinfo{author}{Clarke-Pearson, D.~L.}
\newblock \bibinfo{journal}{\bibinfo{title}{Comparing the areas under two or more correlated receiver operating characteristic curves: a nonparametric approach.}}
\newblock {\emph{\JournalTitle{Biometrics}}} \textbf{\bibinfo{volume}{44}}, \bibinfo{pages}{837--845} (\bibinfo{year}{1988}).

\bibitem{diffusionBeatGANs}
\bibinfo{author}{Dhariwal, P.} \& \bibinfo{author}{Nichol, A.}
\newblock \bibinfo{journal}{\bibinfo{title}{Diffusion models beat gans on image synthesis}}.
\newblock {\emph{\JournalTitle{Advances in neural information processing systems}}} \textbf{\bibinfo{volume}{34}}, \bibinfo{pages}{8780--8794} (\bibinfo{year}{2021}).

\bibitem{FVD}
\bibinfo{author}{Salimans, T.} \emph{et~al.}
\newblock \bibinfo{journal}{\bibinfo{title}{Improved techniques for training gans}}.
\newblock {\emph{\JournalTitle{Advances in neural information processing systems}}} \textbf{\bibinfo{volume}{29}} (\bibinfo{year}{2016}).

\bibitem{I3Dmodel}
\bibinfo{author}{Yang, Z.}, \bibinfo{author}{An, G.}, \bibinfo{author}{Zhang, R.}, \bibinfo{author}{Zheng, Z.} \& \bibinfo{author}{Ruan, Q.}
\newblock \bibinfo{journal}{\bibinfo{title}{Sri3d: Two‐stream inflated 3d convnet based on sparse regularization for action recognition}}.
\newblock {\emph{\JournalTitle{IET Image Processing}}} \textbf{\bibinfo{volume}{17}}, \bibinfo{pages}{n/a--n/a}, \doiprefix\url{10.1049/ipr2.12725} (\bibinfo{year}{2022}).

\bibitem{PSNR}
\bibinfo{author}{Horé, A.} \& \bibinfo{author}{Ziou, D.}
\newblock \bibinfo{title}{Image quality metrics: Psnr vs. ssim}.
\newblock In \emph{\bibinfo{booktitle}{2010 20th International Conference on Pattern Recognition}}, \bibinfo{pages}{2366--2369}, \doiprefix\url{10.1109/ICPR.2010.579} (\bibinfo{year}{2010}).

\bibitem{SSIM}
\bibinfo{author}{Wang, Z.}, \bibinfo{author}{Bovik, A.}, \bibinfo{author}{Sheikh, H.} \& \bibinfo{author}{Simoncelli, E.}
\newblock \bibinfo{journal}{\bibinfo{title}{Image quality assessment: from error visibility to structural similarity}}.
\newblock {\emph{\JournalTitle{IEEE Transactions on Image Processing}}} \textbf{\bibinfo{volume}{13}}, \bibinfo{pages}{600--612}, \doiprefix\url{10.1109/TIP.2003.819861} (\bibinfo{year}{2004}).

\bibitem{LPIPS}
\bibinfo{author}{Zhang, R.}, \bibinfo{author}{Isola, P.}, \bibinfo{author}{Efros, A.~A.}, \bibinfo{author}{Shechtman, E.} \& \bibinfo{author}{Wang, O.}
\newblock \bibinfo{title}{The unreasonable effectiveness of deep features as a perceptual metric}.
\newblock In \emph{\bibinfo{booktitle}{Proceedings of the IEEE conference on computer vision and pattern recognition}}, \bibinfo{pages}{586--595} (\bibinfo{year}{2018}).

\bibitem{t-SNE}
\bibinfo{author}{van~der Maaten, L.} \& \bibinfo{author}{Hinton, G.~E.}
\newblock \bibinfo{journal}{\bibinfo{title}{Visualizing data using t-sne}}.
\newblock {\emph{\JournalTitle{Journal of Machine Learning Research}}} \textbf{\bibinfo{volume}{9}}, \bibinfo{pages}{2579--2605} (\bibinfo{year}{2008}).

\bibitem{PCA}
\bibinfo{author}{F.R.S., K.~P.}
\newblock \bibinfo{journal}{\bibinfo{title}{Liii. on lines and planes of closest fit to systems of points in space}}.
\newblock {\emph{\JournalTitle{The London, Edinburgh, and Dublin Philosophical Magazine and Journal of Science}}} \textbf{\bibinfo{volume}{2}}, \bibinfo{pages}{559--572}, \doiprefix\url{10.1080/14786440109462720} (\bibinfo{year}{1901}).

\bibitem{LIDC}
\bibinfo{author}{Armato, S. G.~r.} \emph{et~al.}
\newblock \bibinfo{journal}{\bibinfo{title}{The lung image database consortium (lidc) and image database resource initiative (idri): a completed reference database of lung nodules on ct scans.}}
\newblock {\emph{\JournalTitle{Med Phys}}} \textbf{\bibinfo{volume}{38}}, \bibinfo{pages}{915--931} (\bibinfo{year}{2011}).

\bibitem{hu2021lora}
\bibinfo{author}{Hu, E.~J.} \emph{et~al.}
\newblock \bibinfo{title}{Lora: Low-rank adaptation of large language models} (\bibinfo{year}{2021}).
\newblock \eprint{2106.09685}.

\bibitem{zhang2023adding}
\bibinfo{author}{Zhang, L.}, \bibinfo{author}{Rao, A.} \& \bibinfo{author}{Agrawala, M.}
\newblock \bibinfo{title}{Adding conditional control to text-to-image diffusion models}.
\newblock In \emph{\bibinfo{booktitle}{IEEE International Conference on Computer Vision (ICCV)}} (\bibinfo{year}{2023}).

\bibitem{huang2023inspect}
\bibinfo{author}{Huang, S.-C.} \emph{et~al.}
\newblock \bibinfo{journal}{\bibinfo{title}{Inspect: a multimodal dataset for pulmonary embolism diagnosis and prognosis}}.
\newblock {\emph{\JournalTitle{arXiv preprint arXiv:2311.10798}}}  (\bibinfo{year}{2023}).

\bibitem{Grad_Cam}
\bibinfo{author}{Selvaraju, R.~R.} \emph{et~al.}
\newblock \bibinfo{journal}{\bibinfo{title}{Grad-cam: Visual explanations from deep networks via gradient-based localization}}.
\newblock {\emph{\JournalTitle{International Journal of Computer Vision}}} \textbf{\bibinfo{volume}{128}}, \bibinfo{pages}{336–359}, \doiprefix\url{10.1007/s11263-019-01228-7} (\bibinfo{year}{2019}).

\bibitem{captum}
\bibinfo{author}{Kokhlikyan, N.} \emph{et~al.}
\newblock \bibinfo{journal}{\bibinfo{title}{Captum: A unified and generic model interpretability library for pytorch}}.
\newblock {\emph{\JournalTitle{arXiv preprint arXiv:2009.07896}}}  (\bibinfo{year}{2020}).

\bibitem{Image-GPT}
\bibinfo{author}{Chen, M.} \emph{et~al.}
\newblock \bibinfo{title}{Generative pretraining from pixels}.
\newblock In \emph{\bibinfo{booktitle}{International Conference on Machine Learning}} (\bibinfo{year}{2020}).

\bibitem{FRIDADAR2018321}
\bibinfo{author}{Frid-Adar, M.} \emph{et~al.}
\newblock \bibinfo{journal}{\bibinfo{title}{Gan-based synthetic medical image augmentation for increased cnn performance in liver lesion classification}}.
\newblock {\emph{\JournalTitle{Neurocomputing}}} \textbf{\bibinfo{volume}{321}}, \bibinfo{pages}{321--331} (\bibinfo{year}{2018}).

\bibitem{düzyel2023dataaugmentationganincreases}
\bibinfo{author}{Düzyel, O.} \& \bibinfo{author}{Kuntalp, M.}
\newblock \bibinfo{title}{Data augmentation with gan increases the performance of arrhythmia classification for an unbalanced dataset} (\bibinfo{year}{2023}).
\newblock \eprint{2302.13855}.

\bibitem{ding2024data}
\bibinfo{author}{Ding, B.} \emph{et~al.}
\newblock \bibinfo{journal}{\bibinfo{title}{Data augmentation using large language models: Data perspectives, learning paradigms and challenges}}.
\newblock {\emph{\JournalTitle{arXiv preprint arXiv:2403.02990}}}  (\bibinfo{year}{2024}).

\bibitem{wells}
\bibinfo{journal}{\bibinfo{title}{Use of a clinical model for safe management of patients with suspected pulmonary embolism}}.
\newblock {\emph{\JournalTitle{Annals of Internal Medicine}}} \textbf{\bibinfo{volume}{129}}, \bibinfo{pages}{997--1005} (\bibinfo{year}{1998}).

\bibitem{medclip}
\bibinfo{author}{Zhang, S.} \emph{et~al.}
\newblock \bibinfo{journal}{\bibinfo{title}{Large-scale domain-specific pretraining for biomedical vision-language processing}}.
\newblock {\emph{\JournalTitle{arXiv preprint arXiv:2303.00915}}}  (\bibinfo{year}{2023}).

\bibitem{huang2018densely}
\bibinfo{author}{Huang, G.}, \bibinfo{author}{Liu, Z.}, \bibinfo{author}{van~der Maaten, L.} \& \bibinfo{author}{Weinberger, K.~Q.}
\newblock \bibinfo{title}{Densely connected convolutional networks} (\bibinfo{year}{2018}).
\newblock \eprint{1608.06993}.

\bibitem{radimagenet}
\bibinfo{author}{Mei, X.} \emph{et~al.}
\newblock \bibinfo{journal}{\bibinfo{title}{Radimagenet: An open radiologic deep learning research dataset for effective transfer learning.}}
\newblock {\emph{\JournalTitle{Radiol Artif Intell}}} \textbf{\bibinfo{volume}{4}}, \bibinfo{pages}{e210315} (\bibinfo{year}{2022}).

\bibitem{dinov2}
\bibinfo{author}{Oquab, M.} \emph{et~al.}
\newblock \bibinfo{journal}{\bibinfo{title}{Dinov2: Learning robust visual features without supervision}}.
\newblock {\emph{\JournalTitle{arXiv preprint arXiv:2304.07193}}}  (\bibinfo{year}{2023}).

\bibitem{transformer}
\bibinfo{author}{Vaswani, A.} \emph{et~al.}
\newblock \bibinfo{journal}{\bibinfo{title}{Attention is all you need}}.
\newblock {\emph{\JournalTitle{Advances in neural information processing systems}}} \textbf{\bibinfo{volume}{30}} (\bibinfo{year}{2017}).

\bibitem{RSPECT}
\bibinfo{author}{Colak, E.} \emph{et~al.}
\newblock \bibinfo{journal}{\bibinfo{title}{The rsna pulmonary embolism ct dataset.}}
\newblock {\emph{\JournalTitle{Radiol Artif Intell}}} \textbf{\bibinfo{volume}{3}}, \bibinfo{pages}{e200254} (\bibinfo{year}{2021}).

\bibitem{2dvae1}
\bibinfo{author}{Rombach, R.}, \bibinfo{author}{Blattmann, A.}, \bibinfo{author}{Lorenz, D.}, \bibinfo{author}{Esser, P.} \& \bibinfo{author}{Ommer, B.}
\newblock \bibinfo{title}{High-resolution image synthesis with latent diffusion models} (\bibinfo{year}{2021}).
\newblock \eprint{2112.10752}.

\bibitem{2dvae2}
\bibinfo{author}{Blattmann, A.}, \bibinfo{author}{Rombach, R.}, \bibinfo{author}{Oktay, K.} \& \bibinfo{author}{Ommer, B.}
\newblock \bibinfo{title}{Retrieval-augmented diffusion models}, \doiprefix\url{10.48550/ARXIV.2204.11824} (\bibinfo{year}{2022}).

\bibitem{monai-gen}
\bibinfo{author}{Pinaya, W. H.~L.} \emph{et~al.}
\newblock \bibinfo{title}{Generative ai for medical imaging: extending the monai framework} (\bibinfo{year}{2023}).
\newblock \eprint{2307.15208}.

\bibitem{patch-disc}
\bibinfo{author}{Wang, T.-C.} \emph{et~al.}
\newblock \bibinfo{title}{High-resolution image synthesis and semantic manipulation with conditional gans} (\bibinfo{year}{2018}).
\newblock \eprint{1711.11585}.

\bibitem{youden}
\bibinfo{author}{YOUDEN, W.~J.}
\newblock \bibinfo{journal}{\bibinfo{title}{Index for rating diagnostic tests.}}
\newblock {\emph{\JournalTitle{Cancer}}} \textbf{\bibinfo{volume}{3}}, \bibinfo{pages}{32--35}, \doiprefix\url{10.1002/1097-0142(1950)3:1<32::aid-cncr2820030106>3.0.co;2-3} (\bibinfo{year}{1950}).

\bibitem{DiffDRR}
\bibinfo{author}{Gopalakrishnan, V.} \& \bibinfo{author}{Golland, P.}
\newblock \bibinfo{title}{Fast auto-differentiable digitally reconstructed radiographs for solving inverse problems in intraoperative imaging}.
\newblock In \emph{\bibinfo{booktitle}{Workshop on Clinical Image-Based Procedures}}, \bibinfo{pages}{1--11} (\bibinfo{organization}{Springer}, \bibinfo{year}{2022}).

\bibitem{TCIA}
\bibinfo{author}{Clark, K.} \emph{et~al.}
\newblock \bibinfo{journal}{\bibinfo{title}{The cancer imaging archive (tcia): maintaining and operating a public information repository}}.
\newblock {\emph{\JournalTitle{Journal of digital imaging}}} \textbf{\bibinfo{volume}{26}}, \bibinfo{pages}{1045--1057} (\bibinfo{year}{2013}).

\end{thebibliography}

\section*{Declarations}

\bigskip\noindent
\textbf{Data availability.} The CXR-CTPA dataset generated and/or analyzed during the current study are not publicly available due patient privacy regulations but are available from the corresponding author (Noa Cahan - E-mail: noa.cahan@gmail.com ) on reasonable request. The LIDC-IDRI (\url{https://wiki.cancerimagingarchive.net/pages/viewpage.action?pageId=1966254}) dataset is publicly available at the cancer imaging archive (TCIA) \cite{TCIA}.

\bigskip\noindent
\textbf{Code availability.} The code and models generated in this study are available: \url{https://github.com/NoaCahan/X-ray2CTPA}.

\bigskip\noindent
\textbf{Acknowledgments.} We thank Prof. Mayte Suarez-Farinas for her valuable support and guidance during this study. This research was supported by the Israel Science Foundation (ISF), grant no. 20/2629. 

\bigskip\noindent
\textbf{Author contributions statement.} Guarantor of integrity of the entire study, planning and supervision, E.K, G.A, H.G.; data curation or data analysis/interpretation, E.K, Y.B.; study concepts/study design and methodology, N.C, R.G, H.G.; software, visualization and writing of the original draft N.C.; Clinical consulting G.A., E.K.; manuscript drafting or manuscript revision for important intellectual content, all authors.; approval of final version of submitted manuscript, all authors.; agrees to ensure any questions related to the work are appropriately  resolved, all authors.; literature research, N.C., Y.B.; experimental studies, N.C, R.G, H.G.; statistical analysis, N.C., R.G., H.G.; and manuscript editing, R.G, N.C, E.K, E.K, Y.B, H.G.

\bigskip\noindent
\textbf{Conflict of interest/Competing interests.} The authors have no relevant financial or non-financial interests to disclose.



\end{document}